\documentclass[preprint, nopreprintline, authoryear, lefttitle,3p]{elsarticle}

\usepackage{amsmath,amssymb,bbm}
\newcommand{\Rho}{\mathrm{P}}

\usepackage[english]{babel}

\usepackage{hyperref}
\usepackage{color}
\definecolor{dred}{rgb}{0.5,0,0}
\definecolor{dblue}{rgb}{0,0,0.5}
\AtBeginDocument{\hypersetup{%
  hyperindex = {true},
  colorlinks = {true},
  linktocpage = {true},
  plainpages = {false},
  linkcolor = {dblue},
  citecolor = {dblue},
  urlcolor = {dred},
  pdfstartview = {Fit},
  pdfpagemode = {UseOutlines},
  pdfview = {XYZ null null null}
}}

\begin{document}

\title{Cholesky-based multivariate Gaussian regression}

\author[1,2]{Thomas Muschinski\corref{cor1}}
\ead{Thomas.Muschinski@uibk.ac.at}
\author[2]{Georg J. Mayr}
\ead{Georg.Mayr@uibk.ac.at}
\author[1,2]{Thorsten Simon}
\ead{Thorsten.Simon@uibk.ac.at}
\author[1]{Nikolaus Umlauf}
\ead{Nikolaus.Umlauf@uibk.ac.at}
\author[1]{Achim Zeileis}
\ead{Achim.Zeileis@R-project.org}
\address[1]{Faculty of Economics and Statistics, Universit\"at Innsbruck, Austria}
\address[2]{Department of Atmospheric and Cryospheric Sciences, Universit\"at Innsbruck, Austria}
\cortext[cor1]{Corresponding author}

\begin{abstract}

Distributional regression is extended to Gaussian response vectors of dimension
	greater than two by parameterizing the covariance matrix $\Sigma$ of
	the response distribution using the entries of its Cholesky
	decomposition.  The more common variance-correlation parameterization
	limits such regressions to bivariate responses -- higher dimensions
	require complicated constraints among the correlations to ensure
	positive definite $\Sigma$ and a well-defined probability density
	function.  In contrast, Cholesky-based parameterizations ensure
	positive definiteness for all distributional dimensions no matter what
	values the parameters take, enabling estimation and regularization as
	for other distributional regression models.  In cases where components
	of the response vector are assumed to be conditionally independent
	beyond a certain lag $r$, model complexity can be further reduced by
	setting Cholesky parameters beyond this lag to zero a priori.
	Cholesky-based multivariate Gaussian regression is first illustrated
	and assessed on artificial data and subsequently applied to a
	real-world 10-dimensional weather forecasting problem. There the
	regression is used to obtain reliable joint probabilities of
	temperature across ten future times, leveraging temporal correlations
	over the prediction period to obtain more precise and meteorologically
	consistent probabilistic forecasts.
\end{abstract}

\maketitle

\section{Introduction} \label{sec:intro}

Distributional regression models \citep{stasinopoulos2018gamlss} -- also called
generalized additive models for location, scale and shape
\citep[GAMLSS,][]{rigby2005generalized} -- extend generalized additive models
\citep[GAM,][]{hastie1990gam} to allow any parametric distribution for the
response. Each parameter of the response distribution -- not just the mean --
can therefore be linked to an additive predictor.  Many different univariate
response distributions have been employed in such additive distributional
regressions, ranging from zero-inflated and overdispersed count data
\citep{klein2015bayesian, simon2019nwp} to survival analysis
\citep{kohler2017flexible, burke2019flexible} or geoadditive hazards regression
\citep{kneib2007mixed}.

Much fewer applications exist for multivariate response distributions.  A
notable exception is \citep{klein2015} where a bivariate response for childhood
undernutrition in India is modeled with a bivariate Gaussian distribution based
on the two means, variances and the correlation, all with suitable link
functions. However, an extension to higher
dimensions is not straightforward because linking individual pairwise
correlations would not assure that the corresponding prediction of the
covariance matrix $\Sigma$ is positive definite -- which in turn is necessary for
a well-defined probability density function. Moreover, the number of
parameters for $\Sigma$ increases quadratically with the dimension of the
response, thus necessitating some form of regularization for the high model
complexity.

We embed multivariate Gaussian regression into the general distributional
regression or GAMLSS framework by parameterizing $\Sigma$ through the entries
of its basic or modified Cholesky decomposition \citep{poura1999},
respectively.  The resulting parameterizations are unconstrained, meaning that
regardless of the values the additive predictors take, the corresponding
covariance matrix $\Sigma$ is guaranteed to be positive definite. This
facilitates regularization through penalized maximum likelihood or Bayesian
estimation of the regression coefficients because the additive predictors can
be regularized separately.  Furthermore, the Cholesky parameterizations allow
model complexity to be restricted a priori in cases where the response
variables are ordered (for example with respect to time or one dimension in
space).  Namely, a covariance with an $r$-order antedependence structure 
\citep[AD-$r$,][]{gabriel1962, zimmerman1998} can be adopted when a maximum lag
may be assumed for the autocorrelations.

The remainder of this paper is structured as follows: A brief overview of
methods for covariance matrix estimation (without dependence on regressors) in
Sec.~\ref{sec:params} motivates leveraging the basic and modified Cholesky
decompositions of $\Sigma$ for a distributional regression (see
Sec.~\ref{sec:dist_reg}) with a multivariate Gaussian response
(Sec.~\ref{sec:mvnchol}). Multivariate Gaussian regression is first illustrated
and assessed on artificial data in Sec.~\ref{sec:simulation}.  Subsequently, in
Sec.~\ref{sec:application} the model is applied to a ten-dimensional weather
forecasting application and different parameterizations are compared. A
discussion of strengths and limitations of the new Cholesky-based multivariate
Gaussian regression framework is found in Sec.~\ref{sec:discussion}.
Summarizing remarks conclude the paper in Sec.~\ref{sec:conclusions}.

\section{Parameterizations of the covariance matrix} \label{sec:params}

For addressing the challenges in multivariate Gaussian regression described above in
Sec.~\ref{sec:intro}, the important first step is to adopt an unconstrained
parameterization of the covariance matrix $\Sigma$. This not only facilitates
estimation of the parameters with standard optimizers and without complicated
constraints, it also enables different forms of regularizations or restrictions
of the model complexity. Hence, we review different parameterizations of the covariance
matrix proposed in the literature, especially with respect to their suitability in
multivariate Gaussian regression. An overview is provided in Table~\ref{tab:reparam}.

\subsection{Positive definiteness of the covariance matrix}

The covariance $\Sigma$ of a $k$-dimensional random variable $y$ from a multivariate
Gaussian distribution is a symmetric $k \times k$ matrix, containing $k \cdot (k + 1) /2$
unique variances and covariances. However, these parameters cannot be chosen freely when
defining $\Sigma$, but must satisfy 
\begin{equation} z^\top \Sigma \, z > 0 \quad \text{for all} \quad z \neq 0.
\end{equation}
to ensure $\Sigma$ is positive definite. Only in this case does the corresponding
probability density function $f(y \mid \mu, \Sigma)$ exist:
\begin{equation} \label{eq:mvgauss_pdf} 
	f(y \mid \mu, \Sigma) = \frac{1}{\sqrt{(2\pi)^k \mid  \Sigma \mid}}
	\exp\left\{-\frac{1}{2} (y - \mu)^\top \Sigma^{-1} (y-\mu)\right\},
\end{equation}
where $\mu = \mathrm{E}(y)$ is the expectation of $y$.

To ensure positive definiteness, joint restrictions for the elements of $\Sigma$
are necessary. The same is true for two other natural parameterizations, namely
the precision matrix $\Sigma^{-1}$ and the variance-correlation decomposition of
$\Sigma$. Similarly, a parameterization
using the spectral decomposition is interpretable with respect to the eigenstructure
of $\Sigma$, but constraints enter through the orthogonality of the corresponding
eigenvectors \citep{poura2013high}.  
When estimating a fixed covariance matrix from empirical observations, some techniques
ensure positive definiteness -- e.g., glasso \citep{friedman2008sparse} and tapering
\citep{furrer2006covariance} -- while others do not -- e.g., hard thresholding
\citep{bickel2008covariance}.

Ensuring a positive definite $\Sigma$ becomes even more difficult in the context
of a distributional regression -- where parameters underlying the covariance matrix should be
linked to regressor variables.  Here, it is particularly beneficial to employ a
parameterization which ensures positive definiteness without requiring joint
constraints and then to combine this with link functions mapping the
parameters to the real line.  The simplest illustration for this is the case of
a univariate Gaussian distribution (i.e., $k = 1$) with variance $\sigma^2$. To
assure positivity, a log link is typically used, mapping the set of positive
real numbers to an unrestricted predictor \citep{stasinopoulos2018gamlss}.
Another notable case is the bivariate Gaussian distribution (i.e., $k = 2$)
where the variance-correlation decomposition can be adopted with log links for
the two variances and a suitable link for the correlation parameter restricted to the
interval $(-1, 1)$ \citep{klein2015}. It is also possible to extend the
log-link approach to $k > 2$ dimensions by using the matrix logarithm, which maps
positive definite symmetric matrices $\Sigma$ to symmetric matrices $A = \log
\Sigma$ with unconstrained entries \citep{poura2013high}. However, the
disadvantages are that (i)~the parameters in $A$ have no natural interpretation
and (ii)~the matrix logarithm involves a Taylor series expansion that is rather
burdensome to compute.

\begin{table}[t!]
\begin{tabular}{lll}
\hline
Parameterization               & No constraints required  & Natural interpretation \\
                               & for positive definite $\Sigma$ & of parameters          \\
\hline
Covariance                     &                 & \checkmark	          \\
Precision                      &                 & \checkmark             \\
Variance-correlation ($k > 2$) &                 & \checkmark             \\
Spectral decomposition         &                 &                        \\
Matrix logarithm               & \checkmark      &                        \\
Cholesky                       & \checkmark      &                        \\
Modified Cholesky              & \checkmark      & \checkmark             \\
\hline
\end{tabular}
\caption{Possible parameterizations of the covariance matrix in multivariate Gaussian distributions,
along with properties that are crucial for linking the covariance structure to regressor variables in
a distributional regression setup. The (modified) Cholesky decomposition is particularly appealing
as its derivation is less burdensome than the matrix logarithm.}
\label{tab:reparam}
\end{table}

\subsection{Cholesky-based parameterizations} \label{sec:cholesky_cov}

A mathematically and computationally more appealing approach that also yields an
unconstrained parameterization is based on the Cholesky decomposition of $\Sigma$. Any $\Sigma$
can be uniquely decomposed as the product of a positive-diagonal lower triangular matrix $L$ with
its transpose $L^\top$:
\begin{equation} \label{eq:basic_chol} 
\Sigma = L L^{\top} \quad , \quad \Sigma^{-1} = (L^{-1})^{\top} L^{-1}.  
\end{equation}
Subsequently the precision matrix $\Sigma^{-1}$ results from a product based on
the inverse Cholesky factor $L^{-1}$.  Both $L$ and $L^{-1}$ offer
unconstrained parameterizations of $\Sigma$.  Although neither the individual
parameters in $L$ nor those in $L^{-1}$ are easily interpretable, the latter
matrix as a whole has an elegant interpretation.  If $y \sim \mathcal{N}(\mu,
\Sigma)$, multiplication with $L^{-1}$ can be used to uncorrelate $y$: $L^{-1}
(y-\mu) \sim \mathcal{N}(0, I)$.

To obtain parameters that are not only unconstrained but individually interpretable,
\citet{poura1999} suggests a modified Cholesky decomposition that diagonalizes $L^{-1}$:
\begin{equation} \label{eq:mod_chol} 
\Sigma^{-1} = T^{\top} D^{-1} T.
\end{equation}
In setups where the $k$ components of $y$ have a natural order (e.g.,
longitudinal data), the entries of the matrices $T$ and $D$ are related to the
autoregressive structure of $y \sim \mathcal{N}(\mu, \Sigma)$. The elements of
the lower triangular matrix $T$ are denoted $-\phi_{ij}$ ($i < j$) -- where
$\phi_{ij}$ are the coefficients of an autoregression on $y$ -- and the elements of the
diagonal matrix $D$ are denoted as $\psi_i$ ($i = 1, \dots, k$) --
corresponding to the innovation variances:
\begin{eqnarray}
\label{eq:autoreg}
    \hat{y}_j & = & \mu_j + \sum_{i=1}^{j-1} \phi_{ij} \cdot (y_i - \mu_i) \quad
        \text{for} \quad j = 2, \dots, k, \\
\label{eq:innov}
          \psi_i & = & \textrm{var}(y_i - \hat{y}_i) 
	  \quad \text{for} \quad i = 1, \dots, k.
\end{eqnarray}
These intuitive interpretations of the parameters $\phi_{ij}$ and $\psi_i$
facilitate regularization of the high model complexity, particularly when $k$
is large. Suggestions from the literature include: using lasso penalties on
$\phi_{ij}$ \citep{levina2008sparse}; approximating the elements of $T$ and $D$
by low-order polynomials \citep{poura1999, poura2000, pan2017jmcm}; or cutting
off the autocorrelation coefficients at a maximum lag of $r$ \citep{wu2003},
i.e., setting $\phi_{ij}=0$ for higher lags $j-i > r$. The latter approach thus
yields a banded $T$ matrix, corresponding to a so-called order-$r$
antedependence \citep[AD-$r$,][]{gabriel1962, zimmerman1998}. Note that since
$T$ and $L^{-1}$ (Eq.~\ref{eq:basic_chol}) share the same pattern of zeros,
AD-$r$ covariances can be modeled using both the modified and basic Cholesky
parameterizations, although the individual elements of $L^{-1}$ are not
directly interpretable as autocorrelation coefficients.

In summary, both Cholesky-based parameterizations are appealing candidates for
a distributional multivariate Gaussian regression approach. They are relatively
easy to compute, yield an unconstrained parameterization that still ensures
positive definite covariances, can be regularized using frequentist or Bayesian techniques,
and can additionally be restricted to an AD-$r$ antedependence, if the $k$
components are autocorrelated with plausible maximum lag of $r$. The modified
Cholesky decomposition has the advantage that individual parameters are
interpretable while the basic Cholesky is slightly easier to compute.

\section{Distributional regression} \label{sec:dist_reg}

In this section we briefly introduce the general distributional regression framework into 
which we embed Cholesky-based multivariate Gaussian regression in the next
Sec.~\ref{sec:mvnchol}. Specifically, the model specification will be a special case of
the general setup from Sec.~\ref{sec:modelspec} so that the corresponding estimation techniques
-- both frequentist and Bayesian -- from Sec.~\ref{sec:model_estimation} can be leveraged.
The software that can be used to estimate the models is presented in Sec.~\ref{sec:software}.

\subsection{Model specification} \label{sec:modelspec}

The idea in distributional regression
\citep[e.g.,][]{rigby2005generalized, klein2015bayesian, umlauf2018bamlss}
is to adopt some $K$-parametric distribution $\mathbf{\mathcal{D}}$ for the response variable $y$,
linking each of the distributional parameters $\theta_k$, $k = 1, \ldots, K$, to separate
flexible additive predictors $\eta_k$
\begin{equation} \label{eqn:dreg}
y \sim \mathbf{\mathcal{D}}\left(h_{1}(\theta_{1}) = \eta_{1}, \,\,
  h_{2}(\theta_{2}) = \eta_{2}, \dots, \,\, h_{K}(\theta_{K}) =
  \eta_{K}\right),
\end{equation}
typically using known monotonic and twice differentiable link functions $h_k( \cdot )$, mapping the
support of each parameter to the unrestricted real values of the predictors.

The predictors combine additively effects of regressor variable(s) $x_{jk}$, $j = 1, \ldots, J_k$,
with
\begin{equation} \label{eqn:addpred}
\eta_k = f_{1k}(x_{1k}) + \ldots + f_{J_kk}(x_{J_kk}),
\end{equation}
where functions $f_{jk}( \cdot )$ can be, e.g., linear terms,
but also nonlinear effects, varying coefficients, random intercepts,
or spatial effects. Rather than explicitly listing all common types of model terms here, we refer to the
literature on GAM \citep{hastie1990gam, wood2017generalized}, GAMLSS \citep{rigby2005generalized},
or Bayesian versions thereof \citep{umlauf2018bamlss}. In this framework, although functions
$f_{jk}( \cdot )$ may be nonlinear, they can be represented by a linear combination of
so-called basis functions and regression coefficients
$f_{jk}(x_{jk}) = \sum_{l = 1}^{d_{jk}} \beta_{ljk} B_{ljk}(x_{jk})$.
For example, functions $f_{jk}( \cdot )$ could be represented by P-splines
\citep{Eilers+Marx:1996} or thin-plate regression splines \citep{Wood:2003}. Hence,
this representation of functions makes this model class very flexible and well suited for
modeling complex relationships.

\subsection{Model estimation}\label{sec:model_estimation}

In a frequentist setting, distributional regression models are commonly estimated using
Newton-Raphson type algorithms maximizing the (penalized) log-likelihood, where parameter updates are
usually obtained by zig-zag iterations over distribution parameters $\theta_k$ and model terms
$f_{jk}( \cdot )$ (see, e.g., \citealp{rigby2005generalized}). Moreover, to avoid overfitting,
nonlinear terms are estimated using penalization techniques as developed for GAMs
\citep{wood2017generalized}, i.e., the wiggliness of each model term is controlled by separate
smoothing parameters, which can be selected by techniques such as the Akaike information
criterion (AIC). The resulting updating equations are known as penalized iteratively
weighted least squares (IWLS, \citealp{Gamerman:1997}). The great benefit of the generic IWLS
representation using a basis function approach is that in most cases only first and second order
derivatives of the log-likelihood with respect to the predictors are needed to implement a new
distribution. This is taken advantage of in Sec.~\ref{sec:mvnchol} for setting up the estimating
equations for the new Cholesky-based multivariate Gaussian regression model.

In addition to this classical GAM-style penalized estimation, the problem of overfitting can also
be addressed by boosting algorithms developed for distributional regression \citep{mayr2012}
or by Lasso-type penalization including factor fusion \citep{Groll+Hambuckers+Kneib+Umlauf:2019}.

However, in the frequentist framework smoothing parameter optimization for complex
distributional regression models can be problematic and computing valid inferential statistics is 
sometimes difficult or even impossible.
The fully Bayesian approach using Markov chain Monte Carlo (MCMC) simulation techniques is 
particularly attractive in such cases. Here, the model parameters are considered as random rather than 
as fixed, meaning that the parameters $\beta_{ljk}$ in a Bayesian model each follow a prior
distribution and the  estimates are computed using the joint posterior distribution, which is
proportional to the product of likelihood and prior. A common choice is to use multivariate normal
priors for the regression coefficients and inverse Gamma (usually the default for spline based
models) or half-Cauchy priors for smoothing variances (can be advantageous with
random effects) that enforce regularization (inverse smoothing parameter in the frequentist
approach). For details see, e.g., \citet{umlauf2018bamlss}.

For efficiency, MCMC algorithms usually draw parameters from the posterior in blocks from full
conditional distributions, i.e., for each model term $f_{jk}( \cdot )$. The full conditionals are
available in closed form only in rare cases, however, a very efficient approximation can be
constructed by a second order Taylor series expansion of the log-posterior centered at the last
parameter state \citep{Gamerman:1997}, which leads to an IWLS-based Metropolis-Hastings algorithm
with an acceptance step. Thus, also for full Bayesian inference of distributional regression models
only first and second order derivatives are needed. For a more detailed introduction to
Bayesian estimation of distributional regression models see \citet{Umlauf+Kneib:2018}.

\subsection{Software implementation} \label{sec:software}

A general flexible implementation of distributional regression with particular
emphasis on Bayesian estimation is provided in the \textsf{R} package \textbf{bamlss}
\citep{Umlauf+Klein+Simon:2021}. The multivariate Gaussian regression models
with basic and modified Cholesky parameterizations, as introduced in the next section,
are implemented as families for \textbf{bamlss}. For now, these families are made
available in a separate package \textbf{mvnchol}, available from the 
Gitlab server of Universit\"at Innsbruck at \url{https://git.uibk.ac.at/c4031039/mvnchol}.
In the future, we plan to integrate the families into \textbf{bamlss}.

\section{Cholesky-based multivariate Gaussian regression} \label{sec:mvnchol}

This section introduces the novel multivariate Gaussian regression approach we
have developed by blending powerful results from the literature on
Cholesky-based parameterizations with the framework of distributional
regression, briefly reviewed in the previous Sec.~\ref{sec:params}
and~\ref{sec:dist_reg}, respectively. The multivariate Gaussian regression
setup is introduced in Sec.~\ref{sec:mgr} and subsequently combined with either
the basic (Sec.~\ref{sec:basic_chol}) or the modified (Sec.~\ref{sec:mod_chol})
Cholesky parameterization to guarantee a positive definite covariance matrix
$\Sigma$.  In order to leverage the typical strategies for estimation and
regularization of distributional regression models, the log-likelihood of the
multivariate Gaussian regression model is provided in Sec.~\ref{sec:loglik}
along with the first and second derivatives with respect to the predictors.

\subsection{Multivariate Gaussian regression} \label{sec:mgr}

In multivariate Gaussian regression the response $y$ is a length-$k$ vector
assumed to follow a $k$-dimensional Gaussian distribution 
\begin{equation}
  y \sim \mathcal{N}(\mu, \Sigma),
\end{equation}
with probability density function provided in Eq.~\ref{eq:mvgauss_pdf}.

All parameters of $\mathcal{N}$ -- the $k$ components of $\mu$
and the $k \cdot (k + 1) / 2$ parameters specifying $\Sigma$ -- may be linked
to predictors. For the $k$ means in $\mu$ this is straightforward as
these parameters are unconstrained and may take any real value. Therefore,
we simply link them to the corresponding additive flexible predictors
using the identity function.
\begin{equation}
    \mu_i = \eta_{\mu,i}, \quad i = 1,\dots, k.
\end{equation}
In contrast, as already argued in Sec.~\ref{sec:params}, it is not possible
to simply link the $k \cdot (k + 1) / 2$ upper-triangular elements of $\Sigma$
to respective additive predictors. This would not ensure that $\Sigma$ is 
positive definite.

Instead we propose to link either the elements of the basic or modified Cholesky
parameterization of $\Sigma$ to additive predictors. While parameterizations based 
on the Cholesky decomposition have been widely used to estimate fixed covariances 
from sparse observations \citep{poura1999,poura2013high}, we exploit them here 
for estimating covariances that depend on further covariates.

\subsection{Basic Cholesky parameterization}\label{sec:basic_chol}

In the basic Cholesky parameterization, $\Sigma$ is defined through the $k$ diagonal 
elements $\lambda_{ii} > 0$ and $k \cdot (k-1) / 2$ off-diagonal elements
$\lambda_{ij}$ of the inverse Cholesky factor:
\begin{equation} \label{eq:lchol}
        (L^{-1})^\top=
  \begin{pmatrix}
    \lambda_{11} & \lambda_{12} & \lambda_{13} & \cdots & \lambda_{1k} \\
    0 & \lambda_{22} & \lambda_{23} & \cdots & \lambda_{2k} \\
    0 & 0 & \lambda_{33} & \cdots & \lambda_{3k} \\
    \vdots & \vdots & \vdots & \ddots & \vdots \\
    0 & 0 & 0 & \cdots & \lambda_{kk}
  \end{pmatrix},
	\quad \text{where} \quad 
  \Sigma = L L^{\top}.
\end{equation}
Restricting the diagonal elements to be positive ensures a unique decomposition, motivating the
use of a log link on these parameters while the off-diagonal elements may take any real value
so that an identity link can be used:
\begin{eqnarray}
	\log(\lambda_{ii}) & = & \eta_{\lambda,ii}, \quad \text{where} \quad i = 1, \dots, k, \\
	\lambda_{ij} & =&  \eta_{\lambda,ij}, \quad \text{where} \quad
	i = 1, \dots, k-1, \quad \text{and} \quad
	j = i +1, \dots, k.
\end{eqnarray}
Modeling the elements of the inverse Cholesky factor $L^{-1}$ is motivated by the following considerations:
(i)~Unlike for the parameterization based on $L$, no computationally intensive matrix
inversions are required during model estimation. (ii)~There is an autoregressive
interpretation for parameter values equal to zero.

Hence, in some situations, where it is not be necessary to model all $k \cdot (k-1) / 2$ off-diagonal elements,
some elements may be restricted to zero. Namely, when the components
of $y$ have a natural order (e.g., longitudinal data) and a maximum lag in the autocorrelations
is reasonable, then an order-$r$ antedependence \citep[AD-$r$,][]{gabriel1962, zimmerman1998}
model can be employed. This sets all $\lambda_{ij} = 0$ with $j-i > r$.
For large $k$ and small $r$ this yields a significant reduction in model complexity.

\subsection{Modified Cholesky parameterization}\label{sec:mod_chol}

Alternatively, the modified Cholesky decomposition of \citet{poura1999}
diagonalizes the inverse Cholesky factor $(L^{-1})^\top$ from Eq.~\ref{eq:lchol},
yielding $\Sigma^{-1} = T^{\top} D^{-1} T$. The new parameters are
those contained in the diagonal matrix $D$ and the upper unitriangular
$T^\top$.
\begin{equation} \label{eq:dt_chol}
        D =
  \begin{pmatrix}
    \psi_1 & 0 & 0 & \cdots & 0 \\
    0 & \psi_2 & 0 & \cdots & 0 \\
    0 & 0 & \psi_3 & \cdots & 0 \\
    \vdots & \vdots & \vdots & \ddots & \vdots \\
    0 & 0 & 0 & \cdots & \psi_k
  \end{pmatrix},
       \quad 
       T^\top=
  \begin{pmatrix}
    1 & -\phi_{12} & -\phi_{13} & \cdots & -\phi_{1k} \\
    0 & 1 & -\phi_{23} & \cdots & -\phi_{2k} \\
    0 & 0 & 1 & \cdots & -\phi_{3k} \\
    \vdots & \vdots & \vdots & \ddots & \vdots \\
    0 & 0 & 0 & \cdots & 1
  \end{pmatrix}.
\end{equation}
The $\psi_i$ in $D$ and the $\phi_{ij}$ in $T^\top$ are called the innovation
variances and generalized autoregressive parameters of $y$, respectively.  They
have meaningful interpretations when the components of $y$ have a natural order.

Analogously to the basic Cholesky parameterization, a log link is used for the
innovation variances to ensure positive definiteness while the 
generalized autoregressive parameters may take any real values:
\begin{eqnarray}
	\log(\psi_{i}) & = & \eta_{\psi,i}, \quad \text{where} \quad i = 1, \dots, k, \\
	\phi_{ij} & = & \eta_{\phi,ij}, \quad \text{where} \quad
	i = 1, \dots, k-1, \quad \text{and} \quad
	j = i +1, \dots, k.
\end{eqnarray}
Again, it is possible to reduce model complexity when an AD-$r$ model can be
assumed. Similar to the basic Cholesky parameterization, this sets all $\phi_{ij} = 0$
with $j-i > r$.

\subsection{The log-likelihood and its derivatives} \label{sec:loglik}

By rearranging the probability density function of the multivariate Gaussian
distribution (Eq.~\ref{eq:mvgauss_pdf}) we obtain the likelihood of
distributional parameters for an observation vector $y$. For mathematical ease,
we work with the log-transformed likelihood.
\begin{equation}        
        \ell(\mu, L^{-1}|y) = -\frac{k}{2}\log(2\pi) + \log(|L^{-1}|) 
        - \frac{1}{2} (y-\mu)^\top (L^{-1})^\top L^{-1} (y-\mu).
\label{eq:loglik_mat}
\end{equation}
Likelihood-based model estimation maximizes the sum of the individual
log-likelihoods (Eq.~\ref{eq:loglik_mat}) over all $n$ observation vectors
contained in the dataset. For computationally efficient estimation, be it
frequentist or Bayesian, this requires derivatives of the
log-likelihood with respect to the additive predictors. We derive analytical
solutions (\ref{sec:app_bas} and \ref{sec:app_mod}) for the first and second
partial derivatives of $\ell$ with respect to all~$\eta_*$. 
The first derivatives in the basic parameterization are found to be 
\begin{equation} 
  \begin{aligned}
	  \frac{\partial \ell}{\partial \eta_{\mu, i}} &= 
        \sum_{j=1}^k \varsigma_{ij} \tilde{y}_j \\
	    \frac{\partial \ell}{\partial \eta_{\lambda,ii}} &=
        1 - \lambda_{ii} \tilde{y}_i \sum_{m=1}^i (\tilde{y}_m \lambda_{mi}) \\
	  \frac{\partial \ell}{\partial \eta_{\lambda,ij}} &=   
  - \tilde{y}_i \sum_{m = 1}^{j} \left( \tilde{y}_m \lambda_{mj} \right), 
  \end{aligned}
\end{equation}
where $\tilde{y}= y- \mu$ and $\varsigma_{ij} = (\Sigma^{-1})_{ij}$.
The corresponding second derivatives are
\begin{equation}
  \begin{aligned}
	  \frac{\partial^2 \ell}{\partial \eta_{\mu,i}^2} &= - \varsigma_{ii} =
        - \sum_{j = i}^{k} \lambda_{ij}^2 \\
          \frac{\partial^2 \ell}{\partial \eta_{\lambda,ii}^2} &= -2 
	  \lambda_{ii}^2 \tilde{y}_i^2 - \lambda_{ii} \tilde{y}_i \cdot
          \sum_{m=1}^{i-1} (\tilde{y}_m \lambda_{mi}) \\
	   \frac{\partial^2 \ell}{\partial \eta_{\lambda,ij}^2} &= - \tilde{y}_i^2.
  \end{aligned}
\end{equation}
These are always negative, which means likelihood-based estimation of the proposed
regression model is a convex optimization problem.  The same is true for the
modified Cholesky parameterization (\ref{sec:app_mod}).

\section{Simulation study} \label{sec:simulation}

To investigate the finite-sample empirical performance of the novel
multivariate Gaussian regression proposed in Sec.~\ref{sec:mvnchol}, this
section conducts a systematic simulation study with more supplementary results
provided in \ref{sec:appendix_sim}. Specifically, we consider a setup where all
distributional parameters ($\mu$, $\psi$, and $\phi$ from a modified Cholesky
parameterization) of the response variable $y$ depend on a covariate $x$,
either in a linear or nonlinear way (Sec.~\ref{sec:data_gen}). Using a flexible
regression model (Sec.~\ref{sec:spec_reg}) with additive spline-based
predictors (i.e., capable of capturing the true effects) it is investigated how
quickly recovery of the true distributional parameters improves as the sample
size increases (\ref{sec:sim_results}).  These results are supplemented in
\ref{sec:appendix_sim} by investigating model misspecifications and effects of
increasing the dimension of the multivariate response variable.

\subsection{Data generation} \label{sec:data_gen}

\begin{figure}[t!]
\centering
\includegraphics[width = 0.85\textwidth]{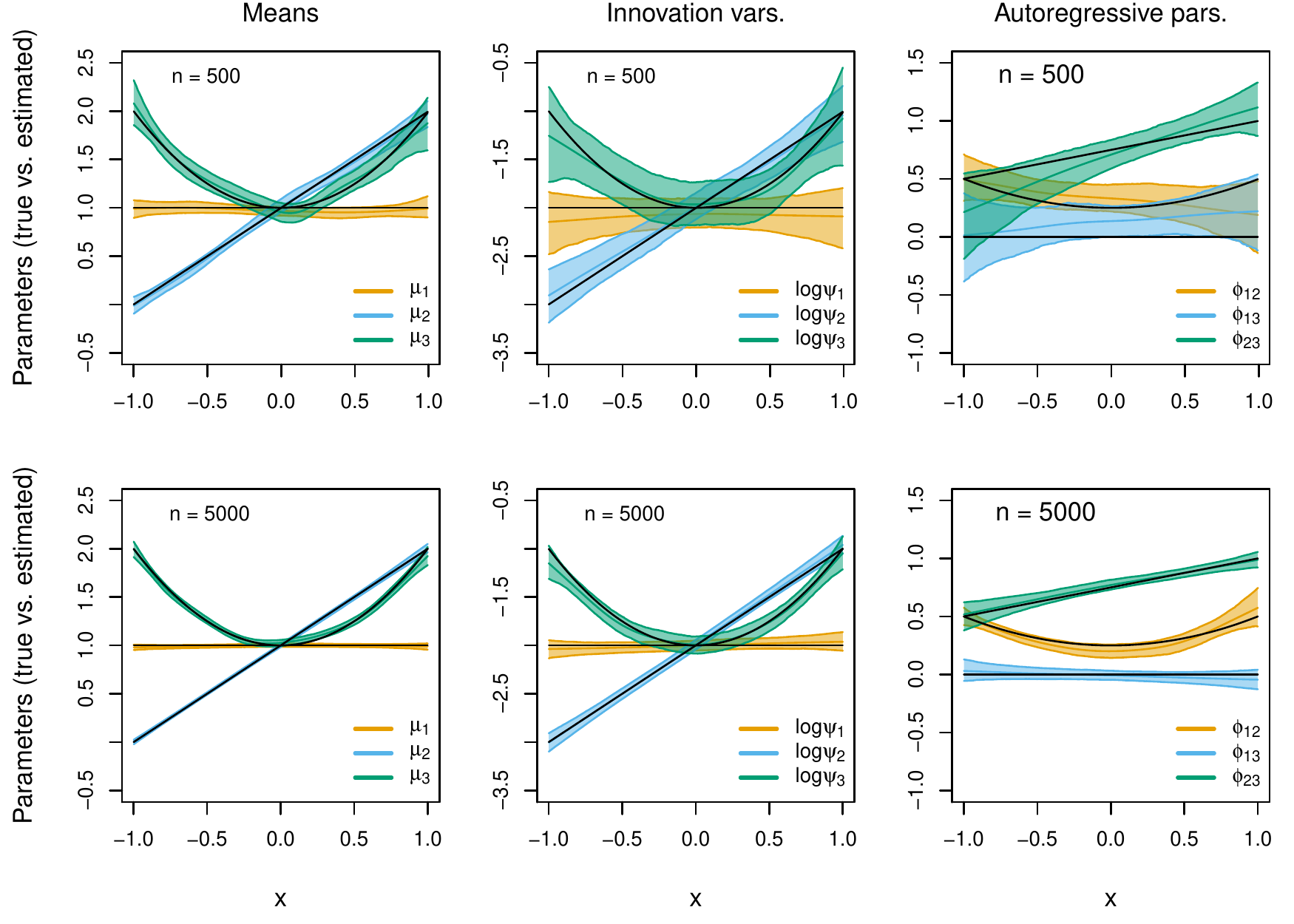}
	\caption{Effects for the three means (left) and modified Cholesky parameters (center and right)
	estimated from datasets of size $n = 500$ (top) and $5000$ (bottom).  Credible intervals
	obtained through MCMC sampling (Fig.~\ref{fig:trace}) are indicated by color shading. The true
	dependencies (Eq.~\ref{data_trivariate}) are depicted by solid black
	lines.}
\label{fig:triv_effects}
\end{figure}

Data sets are constructed by simulating $n$ values of $x$ from a uniform
distribution on the interval $(-1, 1)$. Then for each value of $x$ a 3-dimensional
vector $y = (y_1, y_2, y_3)^\top$ is simulated from a trivariate Gaussian
distribution whose parameters depend on $x$. For each type of parameter
a mixture of constant, linear and quadratic dependencies is used. The exact
equations are given below and visualized by solid black lines in
Fig.~\ref{fig:triv_effects} along with corresponding estimated dependencies
for two simulated data sets of $n = 500$ and $n = 5000$, respectively.
\begin{equation} \label{data_trivariate}
  \begin{aligned}
	  \mu_1 &= 1 \qquad
	  &\log(\psi_{1}) &= -2 \qquad		  
          &\phi_{12} &= (1 + x^2) / 4 \\     
	  \mu_2 &= 1 + x \qquad  
	  &\log(\psi_{2}) &= -2 + x \qquad   
          &\phi_{13} &= 0 \\
	  \mu_3 &= 1 + x^2 \qquad
	  &\log(\psi_{3}) &= -2 + x^2  \qquad
	  &\phi_{23} &= (3 + x) / 4. \\ 
  \end{aligned}
\end{equation}

\begin{figure}[t!]
\centering
\includegraphics[width = 0.85\textwidth]{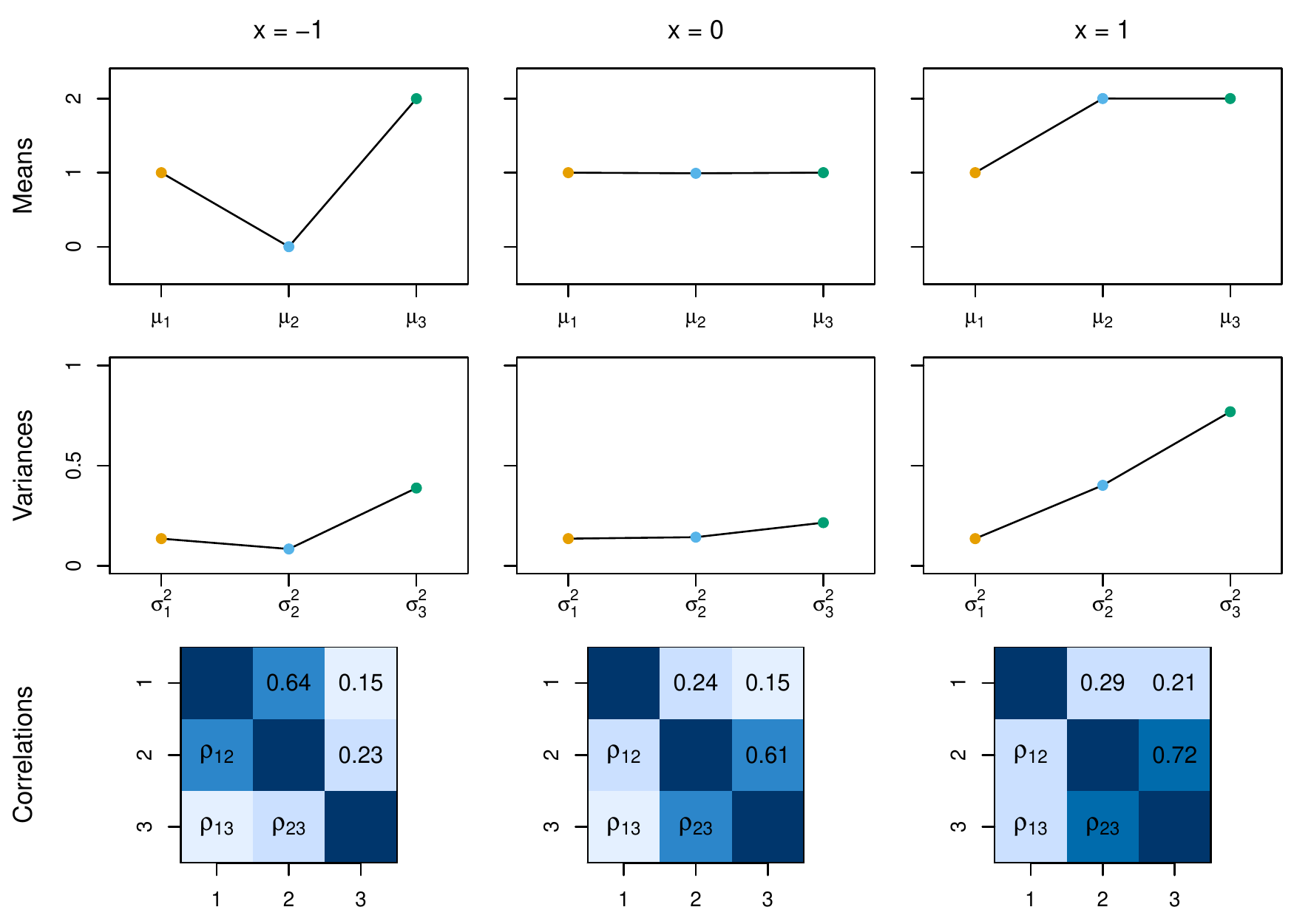} 
	\caption{The means, variances and correlations of the simulated trivariate
	Gaussian distribution (Eq.~\ref{data_trivariate})
	for $x = -1$ (left), $0$ (center), and $1$ (right).}
\label{fig:triv_matrices}
\end{figure}

While Fig.~\ref{fig:triv_effects} emphasizes the dependency of the
distributional parameters (means, innovation variances, and autoregressive
parameters) on the covariate $x$, Fig.~\ref{fig:triv_matrices} brings out how
the corresponding means, variances, and correlations (see Eq.~\ref{eq:dt_chol})
relate across the components of the response $y$.  Three setups are shown,
namely, when computing the parameters for $x = -1$, $0$, and $1$, respectively.

The particular choices for the model specification in Eq.~\ref{data_trivariate}
are made so that the corresponding covariance matrix is of first-order
antedependence (AD-1) type. Specifically, the first variance -- that is always
equal to the first innovation variance -- is kept constant (independent of $x$)
at $\sigma_1^2 = \psi_1 = \exp(-2) \approx 0.14$. Similarly, a constant
$\phi_{13} = 0$ is used so that the first and third components of $y$ are
conditionally independent (i.e., AD-1).  As shown in
Fig.~\ref{fig:triv_matrices}, this does not result in a zero correlation
$\rho_{13}$, but rather one determined by the remaining correlations, i.e.,
$\rho_{13} = \rho_{12} \cdot \rho_{23}$.

\subsection{Regression model specification} \label{sec:spec_reg}

\begin{figure}[t!]
\includegraphics[width = \textwidth]{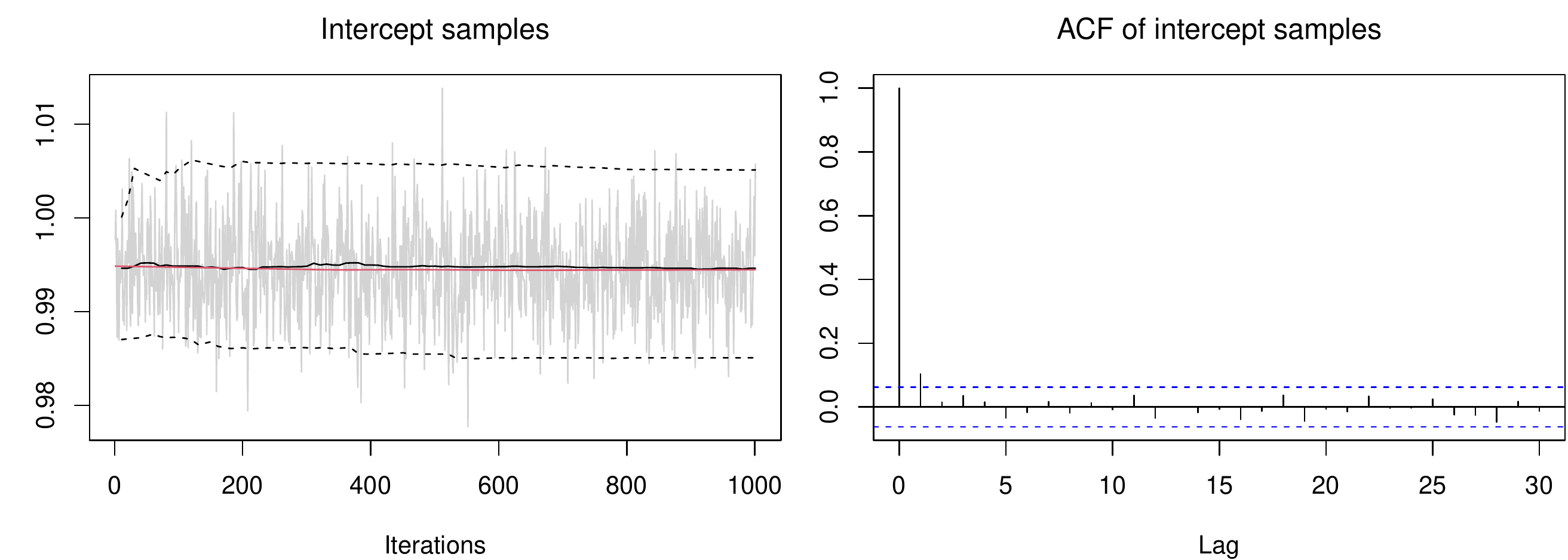}
	\caption{Trace plots of the MCMC samples (left) and autocorrelation
	(right) for the intercept of spline $s_1$ modeling $\mu_1$
	(Eq.~\ref{model_trivariate}).}
\label{fig:trace}
\end{figure}

Multivariate Gaussian regression models employing the modified Cholesky
parameterization are used to estimate the distribution of $y$ conditionally on
$x$.  The three means and six modified Cholesky parameters
are all modeled by thin-plate splines $s_j(x)$ each composed of 10 basis functions:
%
\begin{equation} \label{model_trivariate}
  \begin{aligned}
	  \mu_1 &= s_1(x) \qquad
	  &\log(\psi_{1}) &= s_4(x) \qquad		  
	  &\phi_{12} &= s_7(x) \\     
	  \mu_2 &= s_2(x) \qquad  
	  &\log(\psi_{2}) &= s_5(x) \qquad   
	  &\phi_{13} &= s_8(x) \\
	  \mu_3 &= s_3(x) \qquad
	  &\log(\psi_{3}) &= s_6(x)  \qquad
	  &\phi_{23} &= s_9(x). \\ 
  \end{aligned}
\end{equation}
Bayesian MCMC estimation is employed via the IWLS-based Metropolis-Hastings
algorithm (see Sec.~\ref{sec:dist_reg}). Convergence can be checked using trace
and autocorrelation plots for the regression coefficients $\beta_{ljk}$ from
the spline basis functions (see Sec.~\ref{sec:dist_reg}). Fig.~\ref{fig:trace}
shows these diagnostics for the same simulated data set with $n = 5000$ also
employed in Fig.~\ref{fig:triv_effects}.  Credible intervals for distributional
parameters derived from the MCMC samples are used to test whether or not a
predicted effect is significant.  Overfitting of the complex model is avoided
by choosing prior distributions for the smoothing variances that enforce
regularization of the splines.

\subsection{Results} \label{sec:sim_results}

Fig.~\ref{fig:triv_effects} already conveys that the true distributional
parameters from Eq.~\ref{data_trivariate} are recovered well by the regression
model from Eq.~\ref{model_trivariate}.  The 95\% credible intervals from the
MCMC simulations almost always contain the true values and become much more
narrow as the sample size is increased from $n = 500$ to $5000$. Estimates of
the mean parameters are generally more certain -- i.e., have narrower credible
intervals -- than estimates of the covariance parameters.

\begin{figure}[t!]
\includegraphics[width = \textwidth]{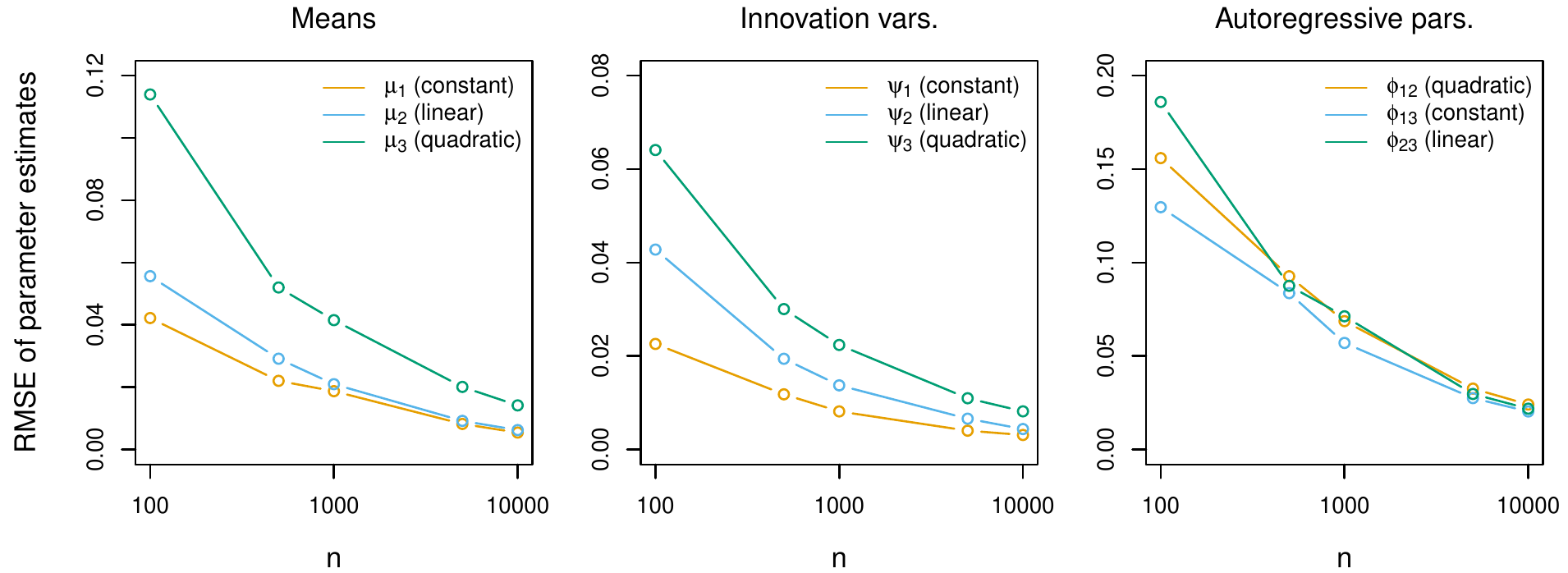}
	\caption{Root-mean-square error (RMSE) of estimates for mean components (left),
	innovation variances (center), and generalized autoregressive parameters
	(right) for increasing sample sizes $n$. True dependencies of the
	parameters on $x$  (i.e., constant, linear or quadratic) are included
	in parenthesis. }
\label{fig:triv_rmse}
\end{figure}

However, the results in Fig.~\ref{fig:triv_effects} are based on only a single
draw for each of the considered sample sizes. To investigate the increasing
predictive skill more thoroughly, we consider 100 replications for each $n =
100, 500, 1000, 5000, 10000$ and assess the root-mean-squared errors (RMSE)
between the true distributional parameters and their corresponding estimates
(see Fig.~\ref{fig:triv_rmse}).  The RMSE is obtained by averaging the errors
at 10000 randomly sampled $x$ from the interval $(-1, 1)$.

For all distributional parameters the RMSE decreases with increasing sample
size~$n$. Also, RMSE increases with increasing complexity of the dependency on
$x$. For a given parameter type (i.e., $\mu$, $\psi$, or $\phi$) constant parameters
are associated with the lowest RMSE, followed by those with linear and quadratic
dependencies.

In addition to these intuitive and reassuring results, \ref{sec:appendix_sim}
shows that virtually the same properties also hold for higher-dimensional responses.
Moreover, it is shown there that the flexible spline specifications are not
very costly in terms of predictive performance. More parsimonious linear specifications
perform only slightly better even when the true effects are linear, but
considerably worse when misspecified.

Finally, to show that the multivariate Gaussian regression model can not only
deal with a single covariate, Sec.~\ref{sec:application} presents a 
10-dimensional weather forecasting problem with 21 covariates.

\section{Application to probabilistic weather forecasting} \label{sec:application}

To illustrate multivariate Gaussian regression in practice, the following multivariate
weather forecasting problem is considered: predicting the temperature for ten future
time points, so-called lead times, simultaneously. For reliable and
meteorologically consistent forecasts, it is crucial to not only accurately
predict the marginal distribution of temperature for each individual time
point, but also the joint distribution of temperature across all times of
interest. This section provides information on numerical weather predictions
(Sec.~\ref{sec:background}), the data construction (Sec.~\ref{sec:GEFS}),
multivariate Gaussian regression specification (Sec.~\ref{sec:nwp_mgr}),
estimated effects and covariance predictions (Sec.~\ref{sec:nwp_effects}),
and evaluation of the model performance (Sec.~\ref{sec:nwp_results}). More
general discussions and comparisons follow in Sec.~\ref{sec:discussion}.

\subsection{Background}
\label{sec:background}

Numerical weather prediction (NWP) models predict the future state of the
atmosphere at multiple lead times by numerically integrating the governing
physical differential equations. The numerical integration begins with a best
guess of the current state of the atmosphere obtained from in-situ and remote
observations around the world. It is typically performed on a discrete grid
approximating the earth-atmosphere system that is several kilometers wide
horizontally and several hundred meters thick vertically \citep{bauer2015quiet}.  

To account for errors from the initialization and
unresolved processes due to the discretization, typically an ensemble of NWP
forecasts is generated using slightly different approximations
\citep{bauer2015quiet,leutbecher2008}. In a final postprocessing step,
statistical regression models are often used for linking actual weather
observations to output from the NWP ensemble outputs in order to improve
forecast accuracy and better calibrate the uncertainty of the predictions
\citep{gneiting2007}. 

Distributional regression has become a popular method to postprocess NWP
ensembles, but its state of the art is limited to univariate \citep[following
the seminal work of][]{gneiting2005calibrated} or bivariate responses
\citep{pinson2012adaptive, schuhen2012ensemble, lang2019bivariate}. However,
some meteorological applications require higher-dimensional joint probability
forecasts across several quantities, locations, or lead times
\citep{feldmann2015spatial, worsnop2018generating, schoenach2020vertical}.
Typically, the prediction errors in such multivariate forecasting problems are
correlated, but rather than estimating this correlation structure as part of
the distributional regression model, it is usually reconstructed from the
empirical NWP ensemble or from historical observations
\citep[e.g.,][]{schefzik2013ecc}.

Here, we leverage the novel Cholesky-based multivariate Gaussian regression model
to predict a full ten-dimensional temperature distribution based on covariates
from an NWP ensemble.

\subsection{Data} \label{sec:GEFS} 

Two-meter temperature forecasts from the Global Ensemble Forecast System
\citep[GEFS,][]{hamill2013noaa} for Innsbruck, Austria, are postprocessed
simultaneously for ten lead times between 186~hours (+7.75~days) and 240~hours
(+10~days). The 11 NWP ensemble members of the GEFS have
spatial resolutions of approximately 70~km and temporal resolutions of 6 hours.
The forecasts are initialized at 00~UTC of 1798 distinct dates over 5~years
and bilinearly interpolated to the spatial coordinates of Innsbruck.

Following \citet{gneiting2005calibrated} and \citet{gebetsberger2019} the
means $\mathtt{mean}_i$ and log-transformed standard deviations $\mathtt{logsd}_i$
of the GEFS ensemble members for each leadtime $i$ are used as covariates
for the statistical postprocessing model. Additionally, the day of the year
$\mathtt{yday}$ of the GEFS run initialization is included to account for
seasonal variations in the postprocessing. See Tab.~\ref{tab:predictor_vars}
for an overview.

\begin{table}[t!]
\begin{tabular}{ll}
\hline
Variable & Description \\
\hline
$\mathtt{mean}_i$ & Ensemble mean temperature forecast for lead time $i$ \\
$\mathtt{logsd}_i$ & Logarithm of ensemble standard deviation for lead time $i$ \\
$\mathtt{yday}$ & Day of year (to capture seasonal variations) \\ 
\hline
\end{tabular} 
\caption{Covariates used as predictor variables to model
	multivariate Gaussian parameters for postprocessing application.
	The placeholder $i$ stands for one of ten lead times (+7.75d, +8d,
	$\dots$, +10d).} 
 \label{tab:predictor_vars}
\end{table} 

Each row of the resulting dataset is associated with a single initialization of the
GEFS on one specific date. There are 10~lead times of interest, so it contains
10~ensemble means and 10~log-transformed standard deviations of the ensemble
temperatures. The 10-dimenstional response variable is composed of the corresponding
observed temperatures $\mathtt{obs}_i$ from the weather station at Innsbruck Airport. 
Since the lead times are spaced 6 hours apart and there is a model initialization
for every day, some observations appear in the response vector for multiple
initializations but as different components. This is not the case for predictors
because NWP forecasts always change from one model initialization to the next.

\subsection{Model specifications} \label{sec:nwp_mgr}

The observed temperature $\mathtt{obs}$ at Innsbruck is modeled for ten
sequential lead times -- every 6 hours between 7.75 and 10 days in the future
-- by a 10-dimensional Gaussian distribution
\begin{equation} 
  y = (\mathtt{obs}_{+7.75d}, \dots, \mathtt{obs}_{+10d})^\top \sim \mathcal{N}(\mu, \Sigma).  
\end{equation}
Distributional parameters of $\mathcal{N}$ are linked to flexible additive
predictors containing $\mathtt{mean}_i$, $\mathtt{logsd}_i$ and
$\mathtt{yday}$.
In all regressions, the ten mean parameters are modeled in the same way:
\begin{equation} \label{model_means}
  \mu_i = s_{0,i}(\mathtt{yday}) + 
          s_{1,i}(\mathtt{yday}) \cdot \mathtt{mean}_i.
\end{equation}
These are linear models of the ensemble mean forecasts, but with seasonally
varying intercepts and slopes estimated by nonlinear cyclical splines $s_{0,i}$
and $s_{1,i}$, respectively.  The regressions differ in how the covariance
matrix is parameterized -- based on its Cholesky or variance-correlation
decomposition -- and subsequently how flexibly it may be modeled (Tab.~\ref{tab:models}). 

\begin{table}
	\begin{tabular}{p{5.4cm}lrrr}
\hline
Model name & Section & \multicolumn{3}{l}{No.\ of covariance parameters} \\
           &         & Flexible & Intercept & Zero \\
\hline
\textit{Basic Cholesky} & \ref{sec:nwp_chol_unstr} & 55 & 0 & 0 \\
\textit{Modified Cholesky} & \ref{sec:nwp_chol_unstr} & 55 & 0 & 0 \\
\textit{Basic Cholesky AD5} & \ref{sec:nwp_chol_ad5} & 45 & 0 & 10 \\
\textit{Modified Cholesky AD5} & \ref{sec:nwp_chol_ad5} & 45 & 0 & 10\\
\textit{AR1} & \ref{sec:nwp_reference} & 11 & 0 & 44 \\
\textit{Constant correlation} & \ref{sec:nwp_reference} & 10 & 45 & 0 \\
\hline
\end{tabular} 
\caption{A 10-dimensional error covariance matrix $\Sigma$ has 55 degrees of
	freedom.  The regression models compared employ different
	parameterizations of $\Sigma$. Depending on the parameterization,
	parameters may either be modeled on predictors, estimated as intercepts
	or are restricted to zero a priori.} 
 \label{tab:models}
\end{table}

\subsubsection{Cholesky parameterizations with fully flexible $\Sigma$} \label{sec:nwp_chol_unstr}
Both the basic and modified Cholesky parameterizations permit all 55
covariance-specifying parameters to be linked to covariates.  The \textit{modified Cholesky}
parameterization employs the following setup:
\begin{equation} \label{model_chol}
  \begin{aligned}
	   \log(\psi_{i}) &= g_{0,i}(\mathtt{yday}) +  
          g_{1,i}(\mathtt{yday}) \cdot \mathtt{logsd}_i\\
	  \phi_{ij} &= h_{ij}(\mathtt{yday}).\\           
  \end{aligned}
\end{equation}
Again, $g_{0,i}$ and $g_{1,i}$ are nonlinear cyclical functions of the year
day, but this time the linear models relate the log-transformed ensemble
standard deviations with the innovation variances $\psi_{i}$.  
Seasonal variations are permitted for the generalized autoregressive parameters 
$\phi_{ij}$ and approximated by cyclical splines $h_{ij}$.

The corresponding \emph{basic Cholesky} parameterization employs the analogous
setup, simply replacing the innovation variances $\psi_{i}$ with $\lambda_{ii}$
and the generalized autoregressive parameters $\phi_{ij}$ with $\lambda_{ij}$.

\subsubsection{Cholesky parameterizations with assumed structure for $\Sigma$}\label{sec:nwp_chol_ad5}
Motivated by a seasonal autoregressive model, an antedependence model of order
5 (AD-5) may be adopted for the covariance structure. Namely, combining
autocorrelations for lag 1 (previous lead time, 6~hours ago) and lag 4
(previous day, 24~hours ago) in a multiplicative way would lead to
autocorrelations up to lag 5 and these are captured here by the AD-5 specificiation.
Thus, only the innovation variances and autoregressive parameters with lags at most 5 
are modeled for the covariance as in Eq.~\ref{model_chol} and higher lag parameters are fixed
at zero. This model is referred to as \textit{modified Cholesky AD5}:
\begin{equation} \label{model_chol_ad5}
          \phi_{ij} = 
          \begin{cases}
            h_{ij}(\mathtt{yday}), & \text{if}\ j-i \leq 5 \\
            0                        , & \text{if}\ j-i > 5 
          \end{cases}           
\end{equation}
Assuming a covariance of type AD-$r$ has the advantage that the number of
covariance parameters increases linearly with the dimension $k$ rather than
quadratically, as with unstructured covariances.

For the corresponding \emph{basic Cholesky AD5} parameterization, the $\psi_i$
are again replaced by $\lambda_{ii}$ and the $\phi_{ij}$ with $\lambda_{ij}$.

\subsubsection{Reference methods using a variance-correlation parameterization} 
\label{sec:nwp_reference}

The Cholesky-based multivariate Gaussian regression models are compared to two
reference methods which parameterize $\Sigma$ through its standard deviations
$\sigma_i$ and correlations $\rho_{ij}$.  In both reference models, standard
deviations are linked to the same additive predictors as were the diagonal
elements of the basic and modified Cholesky decompositions.  Again a log-link
is required to ensure the estimated parameters are positive:
\begin{equation} \label{eq:model_cor}
          \log(\sigma_{i}) = g_{0,i}(\mathtt{yday}) +  
          g_{1,i}(\mathtt{yday}) \cdot \mathtt{logsd}_i.
\end{equation}

The problem with a variance-correlation parameterization is that positive
definiteness is not generally guaranteed when linking all correlations to
predictors (Sec.~\ref{sec:params}). It is possible though to estimate a model
where the correlation matrix $\Rho$ is assumed to conditionally 
follow a first order autoregressive structure. 
\begin{equation}\label{eq:ar1cor}
 \Rho=
  \begin{pmatrix}
   1 & \rho & \rho^2 & \cdots & \rho^{9} \\
   \rho & 1 & \rho & \cdots & \rho^{8} \\
   \rho^2 & \rho & 1 & \cdots & \rho^{7} \\
   \vdots & \vdots & \vdots & \ddots & \vdots \\
   \rho^{9} & \rho^{8} & \rho^{7} & \cdots & 1
  \end{pmatrix}.
\end{equation}
As a result, $\Rho$ is determined by one single parameter $\rho$ instead of
$k \cdot (k -1) /2$ correlations. $\Sigma$ is positive definite if $|\rho| < 1$, 
which allows us to model seasonally varying $\Rho$, but
comes at the cost of very inflexible assumptions about the covariance
across the $k = 10$ lead times.  This model is referred to as \textit{AR1}
and can be denoted by
\begin{equation} \label{eq:model_ar1}
  r(\rho) = h(\mathtt{yday}),          
\end{equation}
where $r = \rho / \sqrt{1 - \rho ^2}$ is the link function mapping the range 
of the parameter $(-1,1)$ to the unrestricted predictor as for the correlation in 
the bivariate Gaussian regression of \citet{klein2015}.

Finally, we compare the Cholesky-based parameterizations against another
alternative model with \textit{constant correlation} for each element $i$, $j$.
In terms of the distributional regression model this means that every correlation
parameter is modeled as an intercept only.
\begin{equation} \label{eq:model_const}
  r(\rho_{ij}) = \mathrm{intercept}_{ij}.
\end{equation}
Thus, unlike all previous specifications
considered above, the correlation structure remains fixed and does not change
across the days of the year.

\subsection{Estimated effects and predictions}\label{sec:nwp_effects}

\begin{figure}[t!]
  \includegraphics[width = \textwidth]{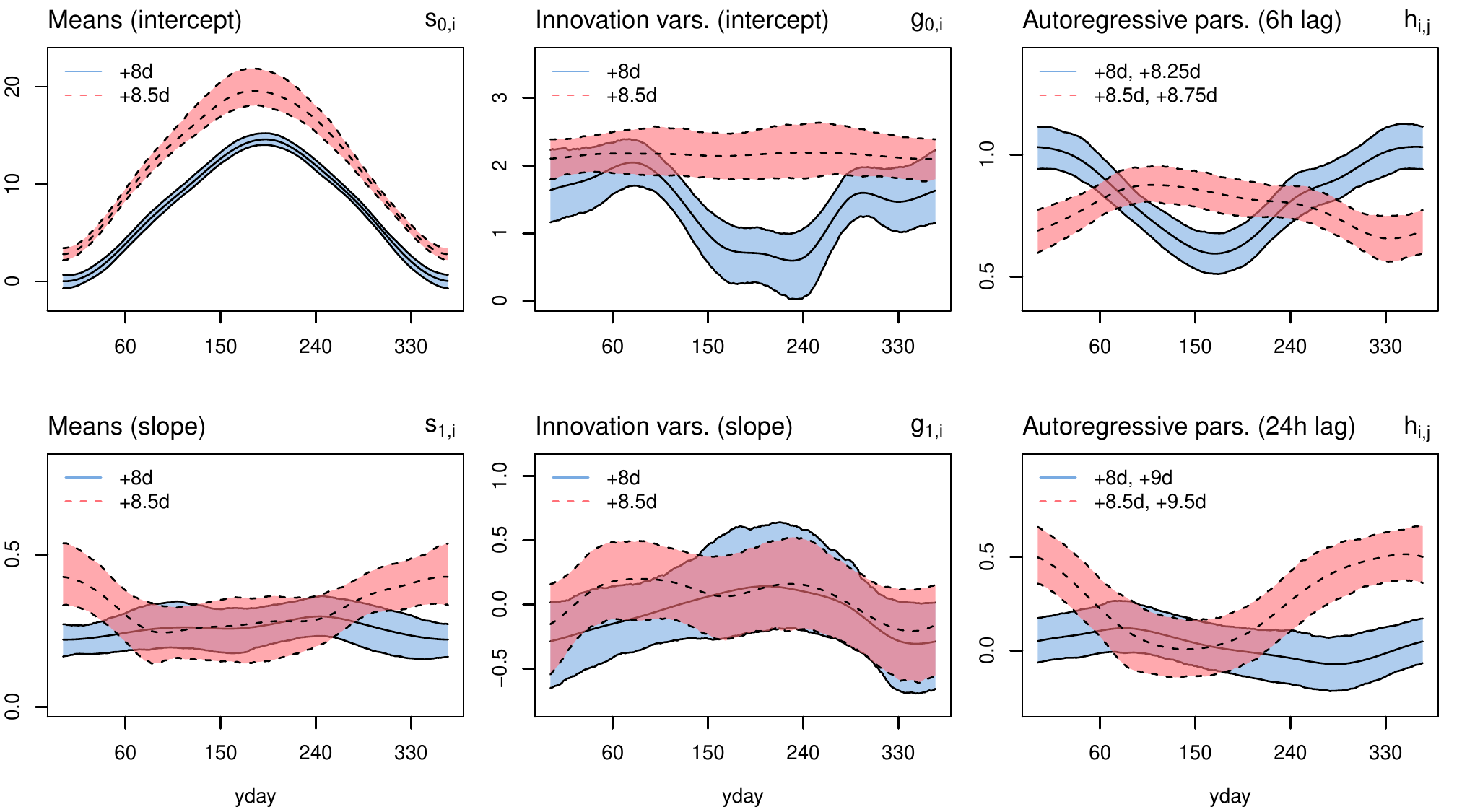}
	\caption{Selected nonlinear effects estimated by the \textit{modified
	Cholesky} model. Shaded regions indicate 95\% credible intervals obtained
	from MCMC sampling. Upper row: Seasonally-varying intercepts for the means (left) and
	log-variances (center) and autoregressive parameters for lag~1 (6h, right).
	Lower row: Seasonally-varying slopes for the means (left) and log-variances (center)
	and autoregressive parameters for lag~4 (24h, right). Red colors indicate forecasts 
	valid for daytime, blue ones for nighttime.}
\label{fig:effects}
\end{figure}

To highlight the flexibility of the Cholesky-based regression models, a
selection of the nonlinear effects estimated by the \textit{modified Cholesky} model are
presented in Fig.~\ref{fig:effects}.
The functions $s_{0,i}$ and $s_{1,i}$ in Eqs.~\ref{model_means}--\ref{model_chol}
can be thought of as seasonally varying intercepts and slopes
in linear models relating ensemble and distributional means.  Slopes $s_{1,i}$
are significantly less than the value of~1 expected for a perfect NWP model
(where the ensemble means/variances would directly correspond to the observed
means/variances).  This means that the GEFS forecast $\mathtt{mean}_i$ contains
limited additional information about the true temperature compared to that
inherent in $\mathtt{yday}$. Therefore intercepts $s_{0,i}$ begin to
approximate a temperature climatology, with summer maxima approximately 15
degrees higher than winter minima (Fig.~\ref{fig:effects}).

For the innovation variances, $g_{0,i}$ and $g_{1,i}$ can again be thought of
as seasonally varying intercepts and slopes in linear models. This time
though, they relate the log-transformed standard deviations of the ensemble
$\mathtt{logsd}_i$ to the log-transformed innovation variances $\psi_{i}$.
Since the GEFS means did not contain much information about the distributional
means, it comes as no surprise that $\mathtt{logsd}_i$ are even less valuable
predictors. The slopes $g_{1,i}$ average close to zero throughout the
year.  Intercepts $g_{0,i}$ have significant seasonal variations for nighttimes (blue)
but not for daytime (red).

The effects $h_{ij} = \phi_{ij}$ can be directly interpreted as seasonal
variations of the generalized autoregressive parameters and paint a complex
picture. For some combinations of $i$ and $j$ the estimated seasonal variations
are significant and for others they are not, with no simple dependency on lag
$j-i$ or index $i$.  

\begin{table}[t!]
  \begin{tabular}{lrrrrr}
  \hline
  Model name & \multicolumn{5}{r}{Runtime for cross validation} \\
             & 1 & 2 & 3 & 4 & 5  \\
  \hline
  \textit{Basic Cholesky} & 118 & 100 & 90 & 75 & 89 \\
  \textit{Modified Cholesky} & 118 & 95 & 166 & 96 & 118 \\
  \textit{Basic Cholesky AD5} & 75 & 79 & 72 & 61 & 66 \\
  \textit{Modified Cholesky AD5} & 102 & 82 & 74 & 73 & 83 \\
  \textit{AR1} & 23 & 23 & 25 & 25 & 20 \\
  \textit{Constant correlation} & 149 & 152 & 151 & 147 & 151 \\
  \hline
  \end{tabular} 
  \caption{Runtime (in minutes) of 10-dimensional Gaussian regression models for
    temperature forecasting.} 
  \label{tab:runtime_nwp}
\end{table} 

\begin{figure}[t!]
\includegraphics[width = 0.95\textwidth]{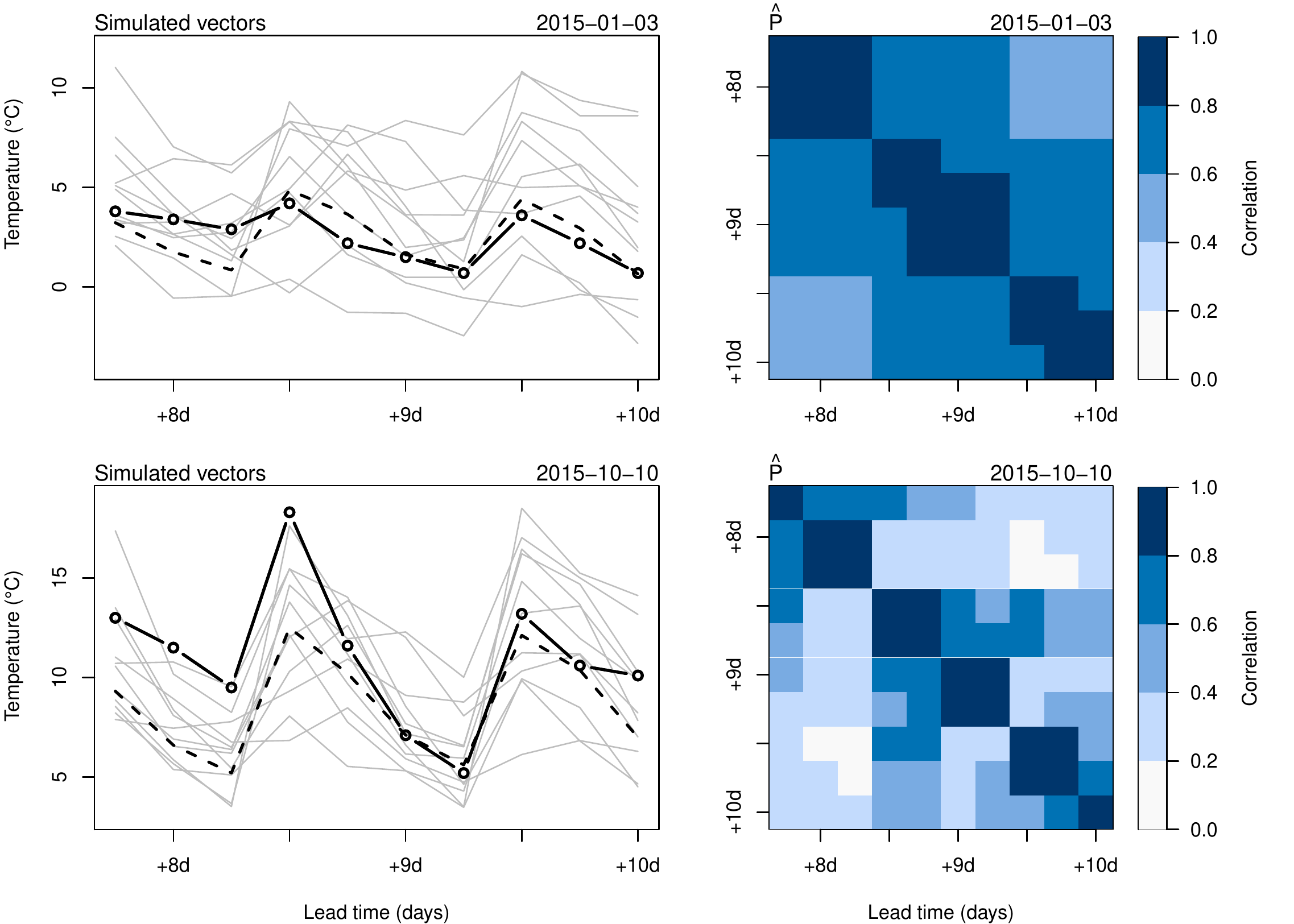}
	\caption{Predictions from the \textit{modified Cholesky} model for $\mu$ and
	$\Sigma$ given values for $\mathtt{yday}$, $\mathtt{mean}$, and $\mathtt{logsd}$
	for two specific days: 2015-01-03 (in winter, top) and 2015-10-10 (in fall, bottom).
	Left: Vectors containing forecast scenarios for the ten lead times are depicted 
	by thin grey lines. These are simulated from the predicted 10-dimensional Gaussian 
	distributions. The means of these distributions are shown as dashed black lines. 
	The true observations are thick black circles 
	connected by lines. Right: Heat maps depicting the
	corresponding estimated correlation matrices $\hat{\mathrm{P}}$.}
\label{fig:trajec}
\end{figure}

Once all $s_\star$, $g_\star$ and $h_\star$ have been estimated (see runtimes
in Tab.~\ref{tab:runtime_nwp}), predictions for the mean $\hat{\mu}$ and
covariance $\hat{\Sigma}$ can be computed from the NWP-derived variables
$\mathtt{yday}$, $\mathtt{mean}_\star$ and $\mathtt{logsd}_\star$.
Fig.~\ref{fig:trajec} visualizes forecasts for two days in~2015: one in winter
(top) and one in fall (bottom). In the left panels, simulated temperature
vectors across all ten lead times are shown in gray along with the actual
observations in black. The mean pattern is approximated reasonably well, albeit
with relatively large variance due to the long lead times. The estimated
correlation matrices $\hat{\Rho}$ for the two days are included in the right
panels.

Clearly the correlation is not constant throughout the year -- as assumed in
the \emph{constant correlation} model -- but differs substantially between
winter and fall. For one, correlations are generally higher in winter, but the
pattern of correlations is also much more complex in fall. In winter a
first-order autoregressive process -- as assumed in the \emph{AR1} model --
might fit reasonably well. However, in the fall, this is not the case and
instead there are large diurnal variations in correlations for a given lag. For
example, forecast errors at 6 UTC in the morning (e.g., +8.25d, +9.25d) have
little influence on the subsequent daytime predictions. This is not the case in
wintertime, where correlations are less variable for a given lag.

\subsection{Model performance} \label{sec:nwp_results}

It is evident that Cholesky-based regressions allow $\Sigma$ to be modeled
flexibly based on the additive predictors. Another question is whether this
increased flexibility improves the quality of the postprocessed joint
probability forecasts.  As the true distributions are unknown, the quality of
the predicted distributions is evaluated using the Dawid-Sebastiani score
\citep[DSS,][]{dawid1999coherent, gneiting2007strictly}.  The DSS is a popular
multivariate score in postprocessing and linearly related to the log-likelihood
of the predicted distributional parameters for a given observation vector.
Scores are evaluated out of sample using five-fold cross-validation. 

\begin{figure}[t!]
\centering
\includegraphics[width = 0.85\textwidth]{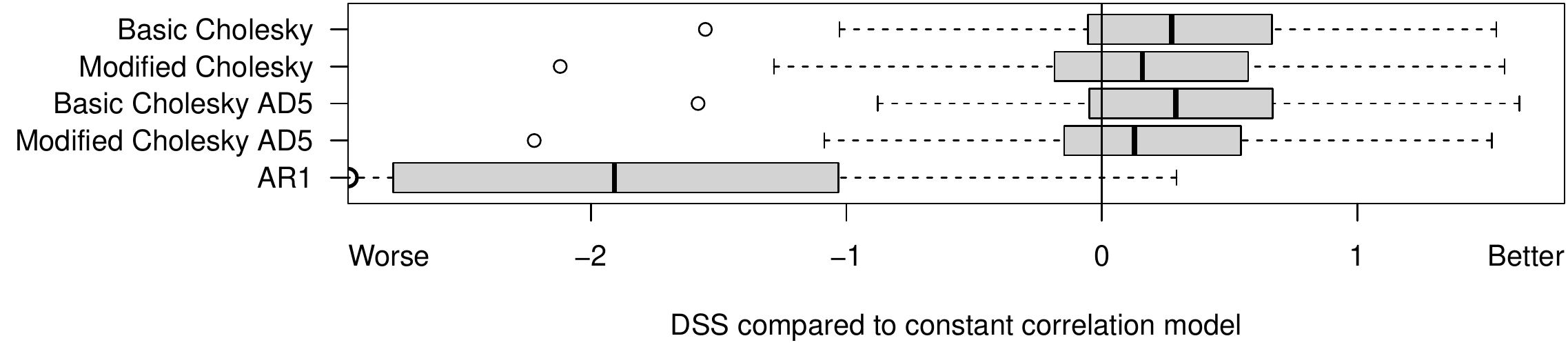}
	\caption{Differences in Dawid-Sebastiani Score (DSS) to the 
	reference \textit{constant correlation} model, aggregated
	by year and month. Positive values mean the given model 
	outperforms the reference. The half circle at the left figure edge 
	indicates that not the entire boxplot for \textit{AR1} is shown
	(minimum value of $-4.6$).}
\label{fig:dss}
\end{figure}

Scores for each method are aggregated by year and month and differences
calculated relative to the reference \emph{constant correlation} model (Fig.\
\ref{fig:dss}).  All Cholesky models perform better than the \textit{constant
correlation} model (vertical line at zero) and much better than the
\textit{AR1} model. The models employing the basic parameterization
(\textit{basic Cholesky} and \textit{basic Cholesky AD5}) are better than the
\textit{constant correlation} model in 75\% of months. The modified Cholesky
models are comparable to the corresponding \emph{basic Cholesky} models, only
very slightly worse.

\section{Discussion} \label{sec:discussion}

The results from the Cholesky-based multivariate Gaussian regression are
discussed further here, in particular regarding the suitability of the novel method
for postprocessing multivariate NWP forecasts (Sec.~\ref{sec:disc_nwp}), its
sensitivity to ordering of the response vector (Sec.~\ref{sec:disc_order}), and
practical limitations to its application along with potential remedies
(Sec.~\ref{sec:disc_future}).

\subsection{Perspectives for multivariate NWP postprocessing} \label{sec:disc_nwp}

In state of the art NWP postprocessing joint probability forecasts
typically do not take the form of joint probability density functions, but are rather 
ensembles obtained through ensemble copula coupling
\citep[ECC,][]{schefzik2013ecc}. In ECC the margins of an NWP ensemble are
calibrated through univariate postprocessing while retaining the ensemble's
order statistics.  

\begin{figure}[t!]
\centering
\includegraphics[width = 0.85\textwidth]{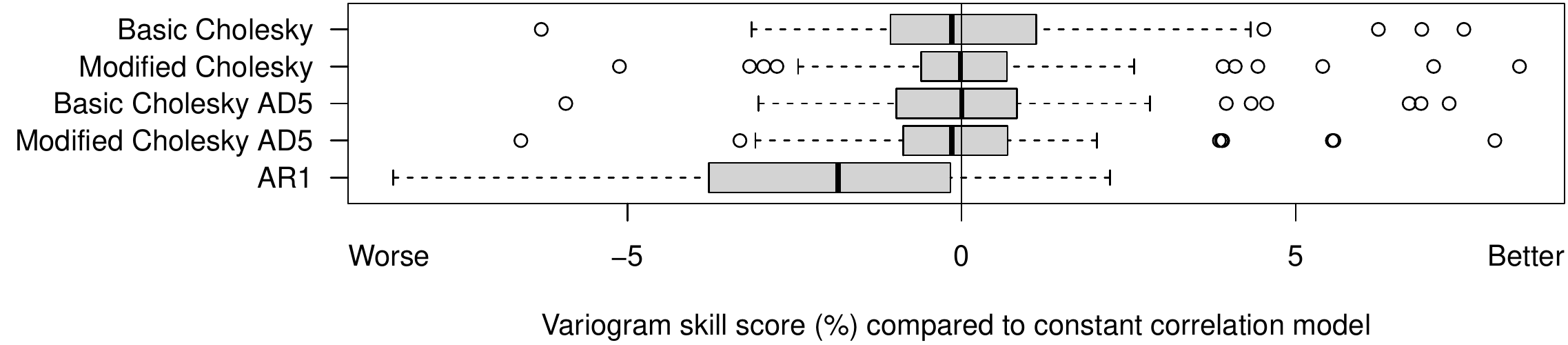}
	\caption{Variogram skill score (\%) relative to the 
	reference \textit{constant correlation} model, based on scores aggregated
	by year and month. Positive values mean the given model 
	outperforms the reference.}
\label{fig:vss}
\end{figure}

In our application, ECC performed much worse than all other
models according to the DSS (Fig.~\ref{fig:dss}, ECC median difference of 6
compared to the \textit{constant correlation} model and not shown). According
to another popular multivariate score -- the variogram score -- ECC again
performed worse than all other postprocessing methods (Fig.~\ref{fig:vss}, ECC
not shown).  Whereas Cholesky-based regression models outperformed all
reference methods according to the DSS, the variogram scores show no
significant improvement compared to the \textit{constant correlation} model.
This may be due to the variogram score being much more sensitive to the
mean and variance than to potential misspecifications of the correlations
\citep{lang2019bivariate}.

The poor performance of ECC is likely due to the overall poor predictive
skill for GEFS forecasts more than a week in advance. ECC may perform more favorably
at shorter lead times, but even here
there is a limit to how well tens of ensemble members can possibly capture
multivariate dependencies with dimensions of the same order. Additionally it is
quite a strong assumption that the ensemble order statistics should reflect
error dependencies across the postprocessed univariate forecasts.
Cholesky-based multivariate Gaussian regression do not rely on these assumptions
and can also be applied when only a direct forecast (and no ensemble) is available.

\subsection{Sensitivity to ordering of the response} \label{sec:disc_order}

\begin{figure}[t!]
\centering
\includegraphics[width = 0.85\textwidth]{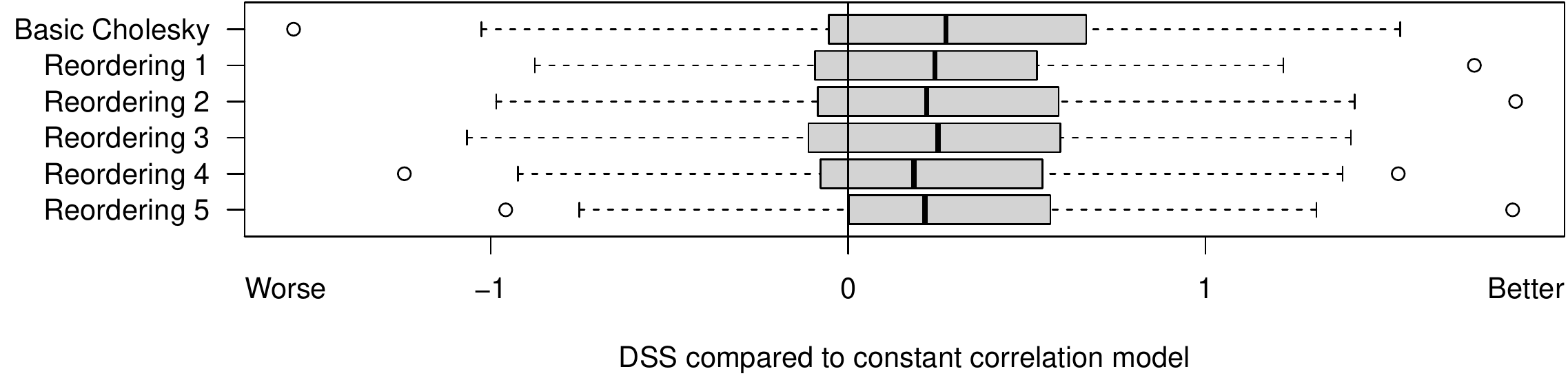}
	\caption{Differences in Dawid-Sebastiani Score (DSS) to \textit{constant correlation}
	as in Fig.~\ref{fig:dss}. The five reorderings have the same 
	model setup as the \textit{basic Cholesky}, but are estimated after random permutations 
	of the variable order.}
	\label{fig:reorder}
\end{figure}

A known limitation of the modified Cholesky decomposition for fixed covariance
estimation is that an ordering of the response components needs to be available
or be assumed \citep{poura2013high}. Many
regularization techniques impose an assumed structure on the parameters which
would be changed by rearranging the components.
Somewhat surprisingly, we find that for our application the
unstructured Cholesky models are quite insensitive to random permutations of
the variables (Fig.~\ref{fig:reorder}). One probable explanation for this is
that the individual regression equations for all distributional parameters
are regularized separately, while in the fixed covariance estimation of
\citet{poura2013high} the ordering is explicitly exploited for imposing
restrictions on the parameters.

\subsection{Future work} \label{sec:disc_future}

Model complexity is still manageable for our 10-dimensional application, but
even here there are 65~distributional parameters and more than 500~model
parameters to estimate from data with a sample size of $n = 1798$ with runtimes
on the order of an hour or two (Tab.~\ref{tab:runtime_nwp}). A fully flexible
parameterization becomes computationally demanding long before $k = 100$, where
5150 distributional parameters would need to be modeled. For very large $k$ it
is also not sufficient to reduce complexity just by assuming $\Sigma$ is
AD-$r$. 

When there is a natural order to the variables, very parsimonious covariance
parameterizations can be obtained by enforcing structure among the Cholesky
parameters. \citet{poura1999} for example assumes polynomial dependencies among
the innovation variances and autoregressive parameters. $\Sigma$ is then
subsequently defined through the coefficients of these polynomials. The
polynomial coefficents could be modeled on predictors in place of the Cholesky
parameters. Alternatively, smooth nonlinear functions may be used to
approximate the parameter structure. Reparameterizations of this sort would
extend the applicability of multivariate Gaussian regression to much higher dimensions.

\section{Conclusions} \label{sec:conclusions}

We have developed regression models for a multivariate Gaussian response, where
all distributional parameters may be linked to flexible additive predictors.
Modeling the mean components of the multivariate dependent variable in such
cases is no different from the univariate case, but it becomes difficult to ensure
the covariance matrix is positive definite for dimensions greater than two.
Common parameterizations such as  variances and a correlation matrix require
joint constraints among parameters to guarantee this property. Such constraints
are difficult to ensure in the context of a regression. 

These challenges are addressed by adopting a parameterization of $\Sigma$ based
on its basic and modified Cholesky decomposition, respectively. These
parameterizations are unconstrained, ensuring positive definite $\Sigma$ for
any predictors.  Subsequently all parameters of the distribution -- the means
and those specifying the covariance -- may be flexibly modeled.  

The ability to model $k \cdot (k + 3) / 2$ distributional parameters comes at
the cost of high complexity. Regression models can be regularized through
penalized likelihood maximization (frequentist approach) or by choosing
appropriate prior distributions (Bayesian). Furthermore, when the
dependent response variable has a natural order, the degrees of freedom of the
covariance matrix may be restricted by assuming a maximum lag for dependencies
among the response components. The triangular matrices in the basic and
modified Cholesky parameterizations of such a covariance are banded.
Subsequently, a large class of parsimonious covariance matrices may be modeled
through a priori restrictions on the parameter space--setting parameters to
zero a priori. This limits model complexity by decreasing the number of
distributional parameters that are linked to predictors.

\begin{appendix}

\section{Basic Cholesky parameterization} \label{sec:app_bas}

The log-likelihood of the distributional parameters for an observation 
vector $y$ is given by
\begin{equation}        
        \ell(\mu, L^{-1}|y) = -\frac{k}{2}\log(2\pi) + \log(|L^{-1}|) 
        - \frac{1}{2} (y-\mu)^\top (L^{-1})^\top L^{-1} (y-\mu).
\label{eq:loglik_mat_app}
\end{equation}
In terms of the individual matrix entries Eq.~\ref{eq:loglik_mat_app} can 
be expressed as
\begin{equation} 
        \ell(\mu, L^{-1}|y) = -\frac{k}{2}\log(2\pi) + \sum_{i=1}^k  
        \log{\lambda_{ii}} - 1/2 z^\top z, 
        \label{eq:ll_chol_terms} 
\end{equation}
where $z$ is the vector
\begin{equation}
        z = L^{-1} \tilde{y}=
  \begin{pmatrix}
    \lambda_{11} & 0 & 0 & \cdots & 0 \\
    \lambda_{12} & \lambda_{22} & 0 & \cdots & 0 \\
    \lambda_{13} & \lambda_{23} & \lambda_{33} & \cdots & 0 \\
    \vdots & \vdots & \vdots & \ddots & \vdots \\
    \lambda_{1k} & \lambda_{2k} & \lambda_{3k} & \cdots & \lambda_{kk}
  \end{pmatrix}
  \begin{pmatrix}
          \tilde{y}_1 \\
          \tilde{y}_2 \\
          \tilde{y}_3 \\
    \vdots \\
          \tilde{y}_k
  \end{pmatrix} \label{eq:define_z}
\end{equation}
and $\tilde{y}= y- \mu$.

The mean parameters and off-diagonal Cholesky entries only influence
the log-likelihood through this third term containing $z$. Partial derivatives with
respect to the means are given by
\begin{equation}
  \frac{\partial \ell}{\partial \mu_i} = 
  \frac{\partial \ell}{\partial \eta_{\mu, i}} = 
        \sum_{j=1}^k \varsigma_{ij} \tilde{y}_j,
  \label{eq:cholscore_mu}
\end{equation}
where $\varsigma_{ij}$ refers to the corresponding
element of $\Sigma^{-1} = (L^{-1})^\top L^{-1}$.

For the off-diagonal Cholesky entries,
\begin{equation} 
   \frac{\partial \ell}{\partial \eta_{\lambda,ij}} = 
   \frac{\partial \ell}{\partial \lambda_{ij}} =  
  -\frac{1}{2} \sum_{n=1}^k \left[2 \left(\sum_{m=1}^n (\tilde{y}_m \lambda_{mn})  
        \right) \tilde{y}_i \mathbbm{1}_j(n) \right] =  
  - \tilde{y}_i \sum_{m = 1}^{j} \left( \tilde{y}_m \lambda_{mj} \right).        
 \label{eq:cholscore_ij} 
\end{equation}
Derivatives with respect to the diagonal entries of $L^{-1}$ also
involve the second likelihood term and are given by
\begin{equation} 
  \frac{\partial \ell}{\partial \lambda_{ii}} =  
        \frac{1}{\lambda_{ii}}- \tilde{y}_i \sum_{m = 1}^{i}  
        \left( \tilde{y}_m \lambda_{mi} \right).         
  \label{eq:cholscore_ii} 
\end{equation}
The log-link on $\lambda_{ii}$ means $\partial \lambda_{ii} / \partial
\eta_{\lambda,ii} = \lambda_{ii}$, and so
\begin{equation}
        \frac{\partial \ell}{\partial \eta_{\lambda,ii}} =
        1 - \lambda_{ii} \tilde{y}_i \sum_{m=1}^i (\tilde{y}_m \lambda_{mi}).
\end{equation}
Second derivatives for parameters with identity link are found to be
\begin{equation}
        \frac{\partial^2 \ell}{\partial \eta_{\mu,i}^2} = - \varsigma_{ii} =
        - \sum_{j = i}^{k} \lambda_{ij}^2,
\end{equation}
\begin{equation}
    \frac{\partial^2 \ell}{\partial \eta_{\lambda,ij}^2} = - \tilde{y}_i^2.
\end{equation}
The log-link on diagonal entries results in
\begin{equation}
  \begin{aligned}
          \frac{\partial^2 \ell}{\partial \eta_{\lambda,ii}^2} &= 
          \frac{\partial}{\partial \eta_{\lambda,ii}} [ 
          1 - \lambda_{ii} \tilde{y}_i \sum_{m=1}^i (\tilde{y}_m \lambda_{mi}) ] \\
	  &= \frac{\partial}{\partial \eta_{\lambda,ii}} [ 1 - \lambda_{ii}^2 
	  \tilde{y}_i^2 ]  -   \frac{\partial}{\partial \eta_{\lambda,ii}} [
		  \lambda_{ii} \tilde{y}_i \sum_{m=1}^{i-1} (\tilde{y}_m \lambda_{mi})] \\
          &=   -2 \lambda_{ii} \cdot \frac{\partial \lambda_{ii}}{\partial \eta_{\lambda,ii}}
          \cdot \tilde{y}_i^2  -  
               \frac{\partial \lambda_{ii}}{\partial \eta_{\lambda,ii}} \cdot 
               \tilde{y}_i \sum_{m=1}^{i-1} (\tilde{y}_m \lambda_{mi}) \\
          &= -2 \lambda_{ii}^2 \tilde{y}_i^2 - \lambda_{ii} \tilde{y}_i \cdot
          \sum_{m=1}^{i-1} (\tilde{y}_m \lambda_{mi}).
  \end{aligned}
\end{equation}

\section{Modified Cholesky parameterization} \label{sec:app_mod}

The modified Cholesky parameters are related to the basic parameters by 
\begin{equation}
  L^{-1} = D^{-1/2} T \qquad \text{which implies} \qquad
        \lambda_{ii} = \psi_i^{-1/2} \quad \text{,} \quad
        \lambda_{ij} = -\phi_{ij} \cdot \psi_j^{-1/2}.
  \label{eq:relate_basic_mod}
\end{equation}
The log-likelihood in Eq.~\ref{eq:ll_chol_terms} can be rewritten
with respect to the new parameters:
\begin{equation}
        \ell(\mu, \psi, \phi | y) = -\frac{k}{2}\log(2\pi) - \frac{1}{2} \sum_{i=1}^k 
        \log{\psi_i} - \frac{1}{2} \sum_{j=1}^k 
        \left( \sum_{i=1}^j \left( \tilde{y}_i \phi_{ij} \psi_j^{-1/2} 
        \right) \right)^2.
 \label{eq:ll_mod_chol}
\end{equation}
For notational simplicity we define $\phi_{ii} = -1$.

The partial derivatives of the log-likelihood with respect to
$\mu$, $\lambda_{ij}$ and $\lambda_{ii}$ (Eqs.~\ref{eq:cholscore_mu},
\ref{eq:cholscore_ij}, \ref{eq:cholscore_ii}) can be related to derivatives 
with respect to the modified Cholesky parameters using Eq.~\ref{eq:relate_basic_mod}:
\begin{equation}
        \frac{\partial \ell}{\partial \mu} = T^\top D^{-1} T \tilde{y},
 \label{eq:mu_modchol}
\end{equation}
\begin{equation}
        \frac{\partial \ell}{\partial \phi_{ij}} = \frac{\partial \ell}{\partial \lambda_{ij}} 
        \cdot \frac{\partial \lambda_{ij}}{\partial \phi_{ij}}
\end{equation}
and
\begin{equation}
        \frac{\partial \ell}{\partial \psi_{i}} = \frac{\partial \ell}{\partial \lambda_{ii}} 
        \cdot \frac{\partial \lambda_{ii}}{\partial \psi_{i}} + 
        \sum_{m=1}^{i-1} \left(  \frac{\partial \ell}{\partial \lambda_{mi}} 
        \cdot \frac{\partial \lambda_{mi}}{\partial \psi_{i}} \right).
\end{equation}
Subsequently
\begin{equation}
        \frac{\partial \lambda_{ij}}{\partial \phi_{ij}} = - \psi_j^{-1/2} 
\end{equation}
and
\begin{equation}
        \frac{\partial \lambda_{ii}}{\partial \psi_{i}} = -\frac{1}{2} \psi_i^{-3/2}
        \qquad \text{,} \qquad
        \frac{\partial \lambda_{ij}}{\partial \psi_{j}} = \frac{1}{2} \phi_{ij} \psi_j^{-3/2}.  
\end{equation}
Substituting the partial derivatives of $\ell$ with respect to $\lambda_{ii}$
and $\lambda_{ij}$ in the basic Cholesky parameterization, one obtains
\begin{equation}
        \frac{\partial \ell}{\partial \phi_{ij}} = - \tilde{y}_i \psi_j^{-1} 
        \sum_{i=1}^j \tilde{y_i} \phi_{ij} 
\end{equation}
and
\begin{equation}
        \frac{\partial \ell}{\partial \psi_{i}} =
        -\frac{1}{2} \left[ \frac{1}{\psi_i} + \tilde{y}_i \sum_{m=1}^i \left(
          \tilde{y}_m \phi_{mi} \psi_i^{-2} \right) \right] + 
        \frac{1}{2} \psi_i^{-2} \sum_{m=1}^{i-1} \left[ \tilde{y}_m \phi_{mi} 
          \left( \sum_{n=1}^i \tilde{y}_n \phi_{ni} \right) \right]
\end{equation}
which simplifies to
\begin{equation}
  \frac{\partial \ell}{\partial \psi_{i}} = \frac{1}{2} \psi_i^{-2} 
    \left[ \left( \sum_{m = 1}^i \tilde{y}_m \phi_{mi} \right) ^2 - \psi_i \right].
\end{equation}
Since $\psi_i$ are estimated using a log-link ($\log(\psi_i) = \eta_{\psi,i}$),
derivatives with respect to predictors become
\begin{equation}           
  \frac{\partial \ell}{\partial \eta_{\psi,i}} =
        \frac{\partial \ell}{\partial \psi_{i}} \psi_{i} 
         =   \frac{1}{2} 
        \left[ \frac{1}{\psi_i} \left( \sum_{m = 1}^i \tilde{y}_m \phi_{mi} \right) ^2 - 1 \right].
\end{equation}
The remaining parameters use an identity link so 
$\partial \ell / \partial \eta_{\phi,ij} = \partial \ell / \partial \phi_{ij}$
and $\partial \ell / \partial \eta_{\mu,i} = \partial \ell / \partial \mu_{i}$.

Continuing with the second derivatives,
\begin{equation}           
        \frac{\partial^2 \ell}{\partial \eta_{\phi,ij}^2} = - \tilde{y}_i^2/ \psi_j 
        \qquad \text{,} \qquad 
        \frac{\partial^2 \ell}{\partial \eta_{\psi,i}^2} = -\frac{1}{2 \psi_i} 
                \left( \sum_{m = 1}^i \tilde{y}_m \phi_{mi} \right) ^2
        \qquad \text{and} \qquad 
        \frac{\partial^2 \ell}{\partial \eta_{\mu,i}^2} = - \varsigma_{ii}, 
\end{equation}
where $\varsigma_{ii}$ is the $i$-th diagonal entry of $\Sigma^{-1}$.

\section{Additional simulations}\label{sec:appendix_sim}

Here, the simulation study of Sec.~\ref{sec:simulation} is extended to test the
influence of nonlinear effects on misspecified linear models
(\ref{sec:appendix_misspec}) as well as model peformance in higher dimensions
(\ref{sec:appendix_dim}). The sample size remains fixed at $n = 5000$.

\subsection{Model misspecifications}\label{sec:appendix_misspec}

The data generating process of Sec.~\ref{sec:simulation} is modified to obtain
datasets with varying degrees of nonlinearity in the parameters of the response
distribution. This is done by multiplying the quadratic terms of $x$ in
Eq.~\ref{data_trivariate} with a nonlinearity parameter $\alpha$:
\begin{equation} \label{data_trivariate_deg}
  \begin{aligned}
	  \mu_1 &= 1 \qquad
	  &\log(\psi_{1}) &= -2 \qquad		  
	  &\phi_{12} &= (1 + \alpha \cdot x^2) / 4 \\     
	  \mu_2 &= 1 + x \qquad  
	  &\log(\psi_{2}) &= -2 + x \qquad   
          &\phi_{13} &= 0 \\
	  \mu_3 &= 1 + \alpha \cdot x^2 \qquad
	  &\log(\psi_{3}) &= -2 + \alpha \cdot x^2  \qquad
	  &\phi_{23} &= (3 + x) / 4. \\ 
  \end{aligned}
\end{equation}
Multivariate Gaussian regression is performed using (i) splines for all
distributional parameters as in Sec.~\ref{sec:simulation} and (ii) linear
models for all distributional parameters.  When $\alpha = 0$, all of the true
dependencies are constant or linear, which means linear models for the
distributional parameters are correctly specified.  When $\alpha$ is increased,
though, the linear models for the parameters with quadratic dependencies on $x$
(i.e., $\mu_3$, $\log \psi_3$, $\phi_{12}$) are misspecified. The dependencies
used in the simulation study of Sec.~\ref{sec:simulation} correspond to $\alpha
= 1$.

\begin{figure}[t!]
\includegraphics[width = \textwidth]{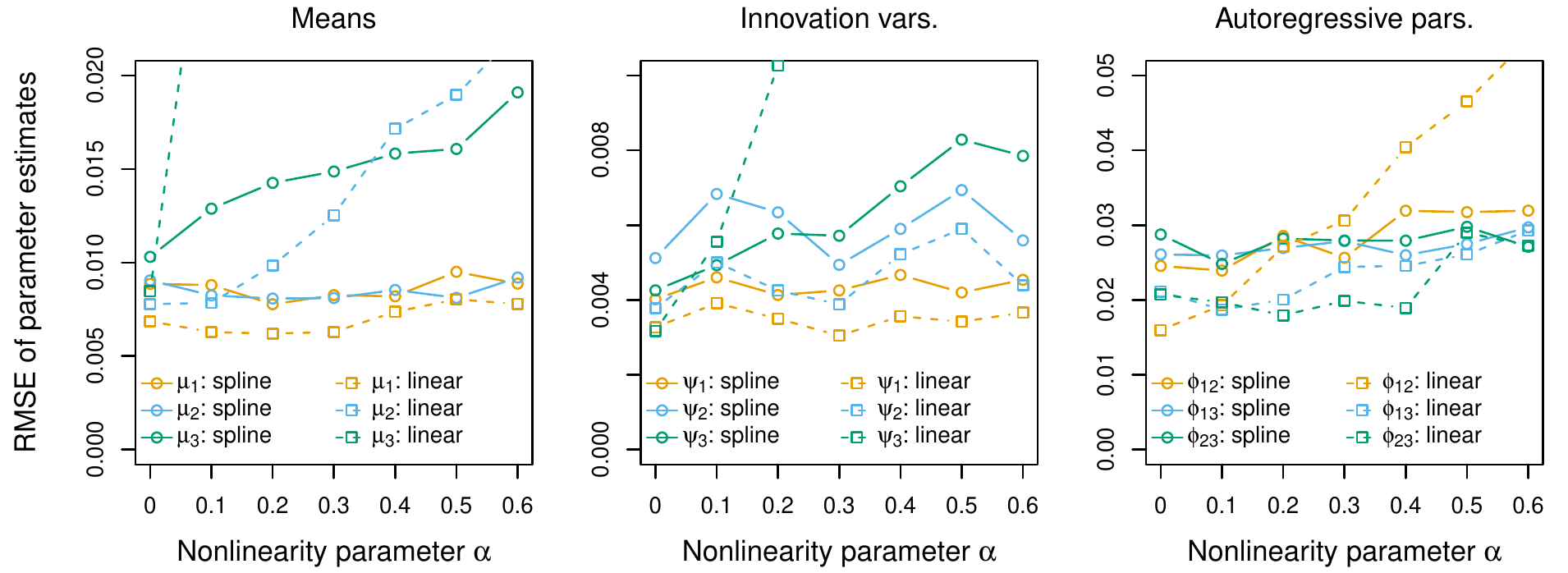}
	\caption{The RMSE of distributional parameter estimates obtained
	using multivariate Gaussian regression with splines (circles, solid lines) 
	or linear models (squares, dashed lines) as a function of the degree 
	of true nonlinearity $\alpha$ (Eq.~\ref{data_trivariate_deg}).}
\label{fig:triv_rmse_deg}
\end{figure}

For the case of true linear dependencies (i.e., $\alpha = 0$), linear
predictors for the distributional parameters perform slightly better than using
splines (Fig.~\ref{fig:triv_rmse_deg}). However, for larger $\alpha$ linear
predictors perform much worse. The RMSE of the misspecified mean parameter
$\mu_3$ triples just by increasing $\alpha$ to 0.1. The increase is more
gradual for the misspecified innovation variance $\psi_3$ and even more so for
the autoregressive parameter $\phi_{12}$. For small $\alpha$ only these terms
perform poorly in the linear specification, but for larger $\alpha$ other terms
deteriorate as well.  Splines in comparison are much more robust to
nonlinearity in the distributional parameter dependencies.

\subsection{Distributional dimension}\label{sec:appendix_dim}

For a trivariate response, multivariate Gaussian regression is already quite
complex. It involves modeling nine distributional parameters by one or more
predictor variables. This complexity increases quadratically
with the dimension.  To investigate how an increase in model complexity
influences the predictive skill of multivariate Gaussian regression models, the
trivariate simulation of Sec.~\ref{sec:simulation} is extended to higher
dimensions $k = 5$, 10 and 15, where there are 20, 65 and 135 distributional
parameters, respectively.

\begin{figure}[t!]
\includegraphics[width = \textwidth]{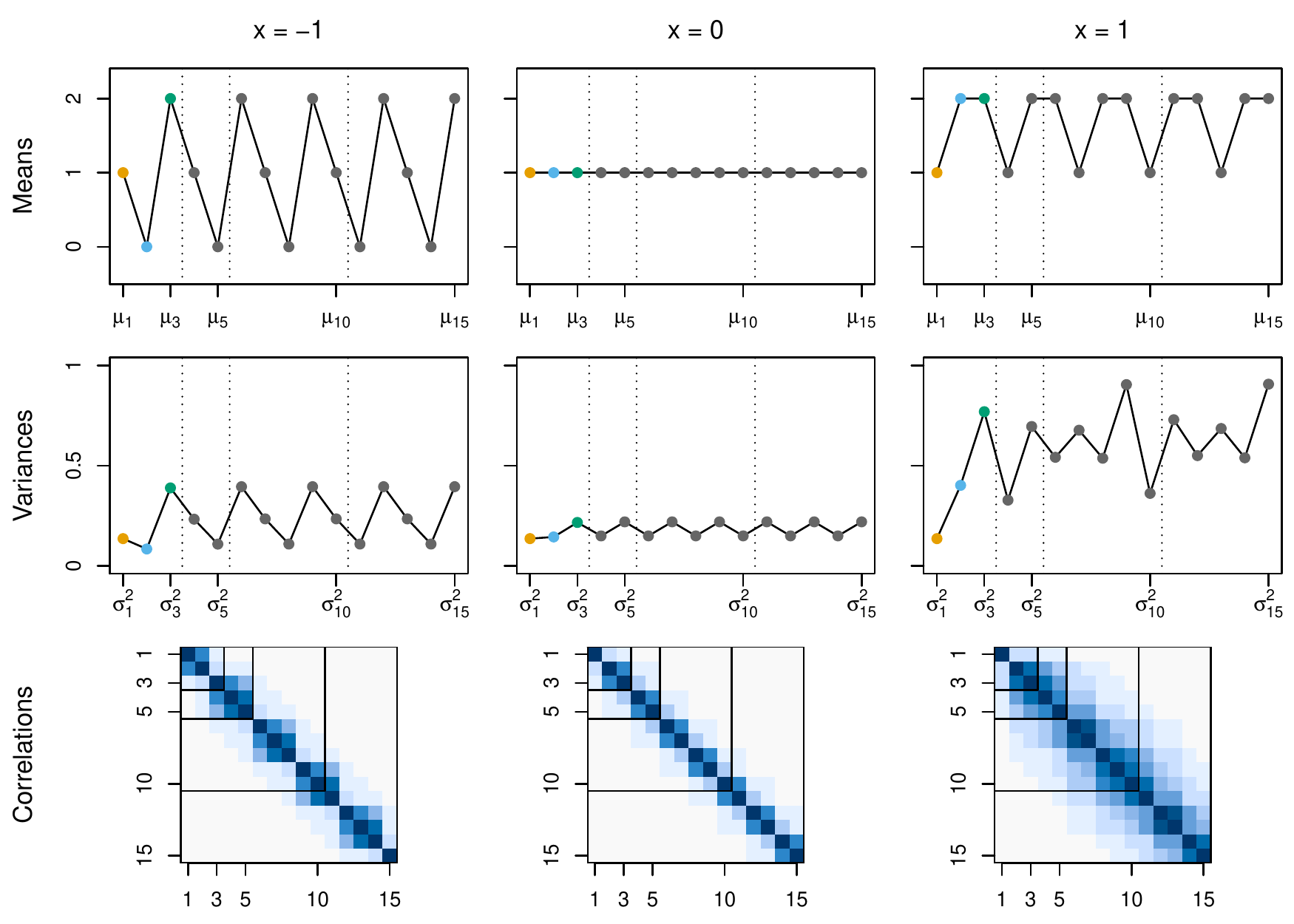}
	\caption{As in Fig.~\ref{fig:triv_matrices}, but extended to include
	means, variances and correlation matrices for the simulating distributions 
	at higher dimensions $k = 5$, 10 and 15. The first three means and 
	variances are colored to match Fig.~\ref{fig:triv_matrices}.}
\label{fig:k_matrices}
\end{figure}

The constant, linear, and quadratic dependencies for the means in
Eq.~\ref{data_trivariate} are repeated so that $\mu_{i+3} = \mu_i$ (i.e.,
$\mu_{11} = \mu_7 = \mu_4 = \mu_1$). Similarly, for the log-transformed
innovation variances $\log \psi_{i+3} = \log \psi_{i}$.  The autoregressive
parameters are slightly different since they have two indices $i$ and $j$.  In
the simulations of Sec.~\ref{sec:simulation}, only the lag-1 autoregressive
parameters $\phi_{12}$ and $\phi_{23}$ depend on $x$; the higher-lag parameter
$\phi_{13}$ is set to 0. This pattern is retained for the higher dimensional
simulations so that $\phi_{ij} = 0$ for all $j - i > 1$. The effects for the
lag-1 parameters $\phi_{12}$ and $\phi_{23}$ are repeated, which results in
$\phi_{(i+2)(i+3)} = \phi_{i(i+1)}$. The means, variances and correlation
matrices of the higher dimensional distributions are visualized for $x = -1$, 0
and -1 in Fig.~\ref{fig:k_matrices} analagously to
Fig.~\ref{fig:triv_matrices}. 

\begin{figure}[t!]
\includegraphics[width = \textwidth]{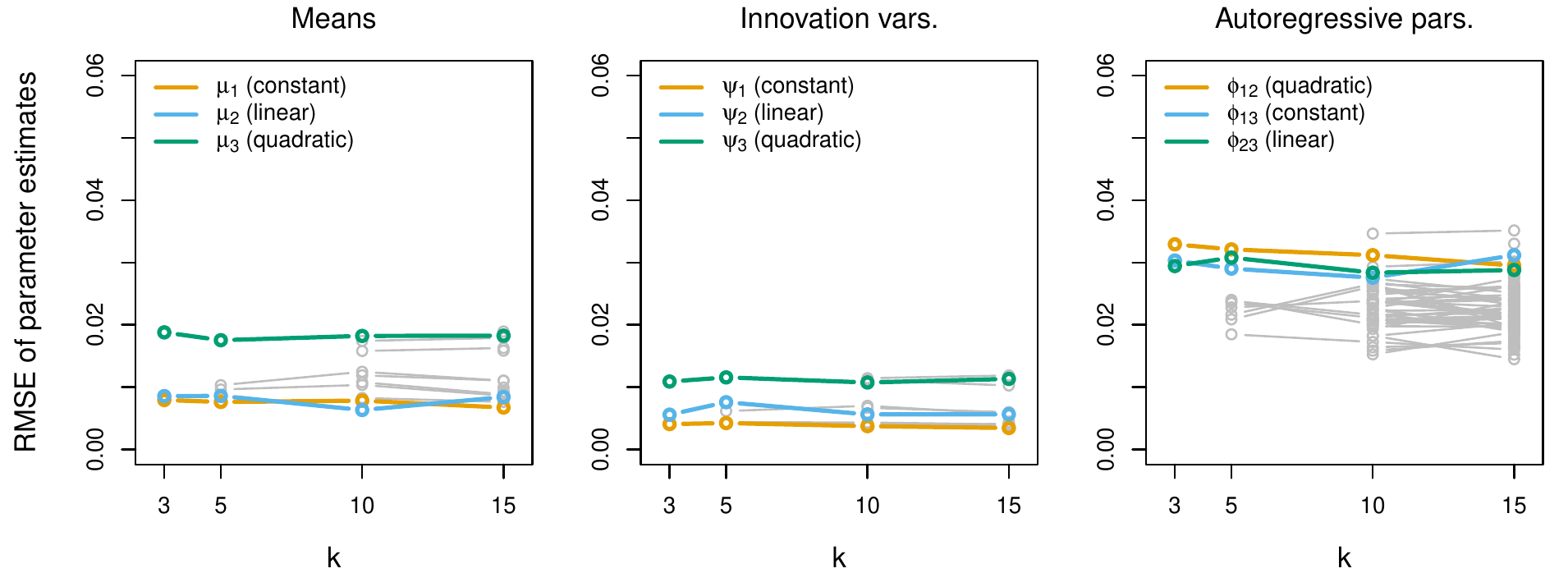}
	\caption{RMSE of estimates for mean components (left column),
	innovation variances (center) and generalized autoregressive parameters
	(right) for different dimensions $k$. Parameters of the trivariate
	distribution are colored and their true dependency on $x$ included in
	parenthesis (i.e., constant, linear or quadratic). Higher dimensional
	parameters are indicated by grey.}
\label{fig:k_rmse}
\end{figure}

The predictive skill of multivariate Gaussian regression models for individual
distributional parameters does not suffer when the dimension is increased
(Fig.~\ref{fig:k_rmse}. The RMSEs for the nine parameters of the trivariate
distribution ($\mu_1, \mu_2, \mu_3, \psi_1, \psi_2, \psi_3, \phi_{12},
\phi_{23}, \phi_{13}$) are nearly identical for $k = 3$, 5, 10 and 15.  The
same is true for higher dimensional means and innovation variances (grey lines
in Fig.~\ref{fig:k_rmse}). For the generalized autoregressive parameters, the
average RMSE is even lower at higher $k$ because a larger fraction of the true
paramters are constants ($\phi_{ij} = 0$ for $j - i > 1$).

\end{appendix}

\section*{Acknowledgements}
This project was funded by the Austrian Science Fund (FWF, grant no.~P\,31836).
The authors thank the Zentralanstalt f\"ur Meteorologie und Geodynamik (ZAMG)
for providing observational data.  The computational results presented here
have been achieved (in part) using the LEO HPC infrastructure of 
Universit\"at Innsbruck. 

\bibliographystyle{elsarticle-harv}
\bibliography{mvnchol}

\begin{thebibliography}{44}
\expandafter\ifx\csname natexlab\endcsname\relax\def\natexlab#1{#1}\fi
\providecommand{\url}[1]{\texttt{#1}}
\providecommand{\href}[2]{#2}
\providecommand{\path}[1]{#1}
\providecommand{\DOIprefix}{doi:}
\providecommand{\ArXivprefix}{arXiv:}
\providecommand{\URLprefix}{URL: }
\providecommand{\Pubmedprefix}{pmid:}
\providecommand{\doi}[1]{\href{http://dx.doi.org/#1}{\path{#1}}}
\providecommand{\Pubmed}[1]{\href{pmid:#1}{\path{#1}}}
\providecommand{\bibinfo}[2]{#2}
\ifx\xfnm\relax \def\xfnm[#1]{\unskip,\space#1}\fi
\bibitem[{Bauer et~al.(2015)Bauer, Thorpe and Brunet}]{bauer2015quiet}
\bibinfo{author}{Bauer, P.}, \bibinfo{author}{Thorpe, A.},
  \bibinfo{author}{Brunet, G.}, \bibinfo{year}{2015}.
\newblock \bibinfo{title}{The quiet revolution of numerical weather
  prediction}.
\newblock \bibinfo{journal}{Nature} \bibinfo{volume}{525}, \bibinfo{pages}{47}.
\newblock \DOIprefix\doi{10.1038/nature14956}.
\bibitem[{Bickel and Levina(2008)}]{bickel2008covariance}
\bibinfo{author}{Bickel, P.J.}, \bibinfo{author}{Levina, E.},
  \bibinfo{year}{2008}.
\newblock \bibinfo{title}{Covariance regularization by thresholding}.
\newblock \bibinfo{journal}{The Annals of Statistics} \bibinfo{volume}{36},
  \bibinfo{pages}{2577--2604}.
\newblock \DOIprefix\doi{10.1214/08-aos600}.
\bibitem[{Burke et~al.(2019)Burke, Jones and Noufaily}]{burke2019flexible}
\bibinfo{author}{Burke, K.}, \bibinfo{author}{Jones, M.C.},
  \bibinfo{author}{Noufaily, A.}, \bibinfo{year}{2019}.
\newblock \bibinfo{title}{A Flexible Parametric Modelling Framework for
  Survival Analysis}.
\newblock \bibinfo{type}{arXiv} \bibinfo{number}{1901.03212}. arXiv.org E-Print
  Archive.
\newblock \URLprefix \url{https://arxiv.org/abs/1901.03212}.
\bibitem[{Dawid and Sebastiani(1999)}]{dawid1999coherent}
\bibinfo{author}{Dawid, A.P.}, \bibinfo{author}{Sebastiani, P.},
  \bibinfo{year}{1999}.
\newblock \bibinfo{title}{Coherent dispersion criteria for optimal experimental
  design}.
\newblock \bibinfo{journal}{The Annals of Statistics} \bibinfo{volume}{27},
  \bibinfo{pages}{65--81}.
\newblock \DOIprefix\doi{10.1214/aos/1018031101}.
\bibitem[{Eilers and Marx(1996)}]{Eilers+Marx:1996}
\bibinfo{author}{Eilers, P.H.C.}, \bibinfo{author}{Marx, B.D.},
  \bibinfo{year}{1996}.
\newblock \bibinfo{title}{Flexible smoothing using {B}-splines and penalized
  likelihood}.
\newblock \bibinfo{journal}{Statistical Science} \bibinfo{volume}{11},
  \bibinfo{pages}{89--121}.
\newblock \DOIprefix\doi{10.1214/ss/1038425655}.
\bibitem[{Feldmann et~al.(2015)Feldmann, Scheuerer and
  Thorarinsdottir}]{feldmann2015spatial}
\bibinfo{author}{Feldmann, K.}, \bibinfo{author}{Scheuerer, M.},
  \bibinfo{author}{Thorarinsdottir, T.L.}, \bibinfo{year}{2015}.
\newblock \bibinfo{title}{Spatial postprocessing of ensemble forecasts for
  temperature using nonhomogeneous {Gaussian} regression}.
\newblock \bibinfo{journal}{Monthly Weather Review} \bibinfo{volume}{143},
  \bibinfo{pages}{955--971}.
\newblock \DOIprefix\doi{10.1175/mwr-d-14-00210.1}.
\bibitem[{Friedman et~al.(2008)Friedman, Hastie and
  Tibshirani}]{friedman2008sparse}
\bibinfo{author}{Friedman, J.}, \bibinfo{author}{Hastie, T.},
  \bibinfo{author}{Tibshirani, R.}, \bibinfo{year}{2008}.
\newblock \bibinfo{title}{Sparse inverse covariance estimation with the
  graphical lasso}.
\newblock \bibinfo{journal}{Biostatistics} \bibinfo{volume}{9},
  \bibinfo{pages}{432--441}.
\newblock \DOIprefix\doi{10.1093/biostatistics/kxm045}.
\bibitem[{Furrer et~al.(2006)Furrer, Genton and Nychka}]{furrer2006covariance}
\bibinfo{author}{Furrer, R.}, \bibinfo{author}{Genton, M.G.},
  \bibinfo{author}{Nychka, D.}, \bibinfo{year}{2006}.
\newblock \bibinfo{title}{Covariance tapering for interpolation of large
  spatial datasets}.
\newblock \bibinfo{journal}{Journal of Computational and Graphical Statistics}
  \bibinfo{volume}{15}, \bibinfo{pages}{502--523}.
\newblock \DOIprefix\doi{10.1198/106186006x132178}.
\bibitem[{Gabriel(1962)}]{gabriel1962}
\bibinfo{author}{Gabriel, K.R.}, \bibinfo{year}{1962}.
\newblock \bibinfo{title}{Ante-dependence analysis of an ordered set of
  variables}.
\newblock \bibinfo{journal}{The Annals of Mathematical Statistics}
  \bibinfo{volume}{33}, \bibinfo{pages}{201--212}.
\newblock \DOIprefix\doi{10.1214/aoms/1177704724}.
\bibitem[{Gamerman(1997)}]{Gamerman:1997}
\bibinfo{author}{Gamerman, D.}, \bibinfo{year}{1997}.
\newblock \bibinfo{title}{Sampling from the posterior distribution in
  generalized linear mixed models}.
\newblock \bibinfo{journal}{Statistics and Computing} \bibinfo{volume}{7},
  \bibinfo{pages}{57--68}.
\newblock \DOIprefix\doi{10.1023/a:1018509429360}.
\bibitem[{Gebetsberger et~al.(2019)Gebetsberger, Stauffer, Mayr and
  Zeileis}]{gebetsberger2019}
\bibinfo{author}{Gebetsberger, M.}, \bibinfo{author}{Stauffer, R.},
  \bibinfo{author}{Mayr, G.J.}, \bibinfo{author}{Zeileis, A.},
  \bibinfo{year}{2019}.
\newblock \bibinfo{title}{Skewed logistic distribution for statistical
  temperature postprocessing in mountainous areas}.
\newblock \bibinfo{journal}{Advances in Statistical Climatology, Meteorology
  and Oceanography} \bibinfo{volume}{5}, \bibinfo{pages}{87--100}.
\newblock \URLprefix \url{https://ascmo.copernicus.org/articles/5/87/2019/},
  \DOIprefix\doi{10.5194/ascmo-5-87-2019}.
\bibitem[{Gneiting et~al.(2007)Gneiting, Balabdaoui and Raftery}]{gneiting2007}
\bibinfo{author}{Gneiting, T.}, \bibinfo{author}{Balabdaoui, F.},
  \bibinfo{author}{Raftery, A.E.}, \bibinfo{year}{2007}.
\newblock \bibinfo{title}{Probabilistic forecasts, calibration and sharpness}.
\newblock \bibinfo{journal}{Journal of the Royal Statistical Society B}
  \bibinfo{volume}{69}, \bibinfo{pages}{243--268}.
\newblock \DOIprefix\doi{10.21236/ada454827}.
\bibitem[{Gneiting and Raftery(2007)}]{gneiting2007strictly}
\bibinfo{author}{Gneiting, T.}, \bibinfo{author}{Raftery, A.E.},
  \bibinfo{year}{2007}.
\newblock \bibinfo{title}{Strictly proper scoring rules, prediction, and
  estimation}.
\newblock \bibinfo{journal}{Journal of the American Statistical Association}
  \bibinfo{volume}{102}, \bibinfo{pages}{359--378}.
\newblock \DOIprefix\doi{10.1198/016214506000001437}.
\bibitem[{Gneiting et~al.(2005)Gneiting, Raftery, Westveld~III and
  Goldman}]{gneiting2005calibrated}
\bibinfo{author}{Gneiting, T.}, \bibinfo{author}{Raftery, A.E.},
  \bibinfo{author}{Westveld~III, A.H.}, \bibinfo{author}{Goldman, T.},
  \bibinfo{year}{2005}.
\newblock \bibinfo{title}{Calibrated probabilistic forecasting using ensemble
  model output statistics and minimum {CRPS} estimation}.
\newblock \bibinfo{journal}{Monthly Weather Review} \bibinfo{volume}{133},
  \bibinfo{pages}{1098--1118}.
\newblock \DOIprefix\doi{10.1175/mwr2904.1}.
\bibitem[{Groll et~al.(2019)Groll, Hambuckers, Kneib and
  Umlauf}]{Groll+Hambuckers+Kneib+Umlauf:2019}
\bibinfo{author}{Groll, A.}, \bibinfo{author}{Hambuckers, J.},
  \bibinfo{author}{Kneib, T.}, \bibinfo{author}{Umlauf, N.},
  \bibinfo{year}{2019}.
\newblock \bibinfo{title}{Lasso-type penalization in the framework of
  generalized additive models for location, scale and shape}.
\newblock \bibinfo{journal}{Computational Statistics \& Data Analysis}
  \bibinfo{volume}{140}, \bibinfo{pages}{59--74}.
\newblock \DOIprefix\doi{10.1016/j.csda.2019.06.005}.
\bibitem[{Hamill et~al.(2013)Hamill, Bates, Whitaker, Murray, Fiorino,
  Galarneau~Jr, Zhu and Lapenta}]{hamill2013noaa}
\bibinfo{author}{Hamill, T.M.}, \bibinfo{author}{Bates, G.T.},
  \bibinfo{author}{Whitaker, J.S.}, \bibinfo{author}{Murray, D.R.},
  \bibinfo{author}{Fiorino, M.}, \bibinfo{author}{Galarneau~Jr, T.J.},
  \bibinfo{author}{Zhu, Y.}, \bibinfo{author}{Lapenta, W.},
  \bibinfo{year}{2013}.
\newblock \bibinfo{title}{{NOAA}'s second-generation global medium-range
  ensemble reforecast dataset}.
\newblock \bibinfo{journal}{Bulletin of the American Meteorological Society}
  \bibinfo{volume}{94}, \bibinfo{pages}{1553--1565}.
\newblock \DOIprefix\doi{10.1175/bams-d-12-00014.1}.
\bibitem[{Hastie and Tibshirani(1990)}]{hastie1990gam}
\bibinfo{author}{Hastie, T.J.}, \bibinfo{author}{Tibshirani, R.J.},
  \bibinfo{year}{1990}.
\newblock \bibinfo{title}{Generalized Additive Models}.
  volume~\bibinfo{volume}{43}.
\newblock \bibinfo{publisher}{Chapman \& Hall/CRC}.
\bibitem[{Klein et~al.(2015a)Klein, Kneib, Klasen and Lang}]{klein2015}
\bibinfo{author}{Klein, N.}, \bibinfo{author}{Kneib, T.},
  \bibinfo{author}{Klasen, S.}, \bibinfo{author}{Lang, S.},
  \bibinfo{year}{2015}a.
\newblock \bibinfo{title}{Bayesian structured additive distributional
  regression for multivariate responses}.
\newblock \bibinfo{journal}{Journal of the Royal Statistical Society C}
  \bibinfo{volume}{64}, \bibinfo{pages}{569--591}.
\newblock \DOIprefix\doi{10.1111/rssc.12090}.
\bibitem[{Klein et~al.(2015b)Klein, Kneib and Lang}]{klein2015bayesian}
\bibinfo{author}{Klein, N.}, \bibinfo{author}{Kneib, T.},
  \bibinfo{author}{Lang, S.}, \bibinfo{year}{2015}b.
\newblock \bibinfo{title}{{B}ayesian generalized additive models for location,
  scale, and shape for zero-inflated and overdispersed count data}.
\newblock \bibinfo{journal}{Journal of the American Statistical Association}
  \bibinfo{volume}{110}, \bibinfo{pages}{405--419}.
\newblock \DOIprefix\doi{10.1080/01621459.2014.912955}.
\bibitem[{Kneib and Fahrmeir(2007)}]{kneib2007mixed}
\bibinfo{author}{Kneib, T.}, \bibinfo{author}{Fahrmeir, L.},
  \bibinfo{year}{2007}.
\newblock \bibinfo{title}{A mixed model approach for geoadditive hazard
  regression}.
\newblock \bibinfo{journal}{Scandinavian Journal of Statistics}
  \bibinfo{volume}{34}, \bibinfo{pages}{207--228}.
\newblock \DOIprefix\doi{10.1111/j.1467-9469.2006.00524.x}.
\bibitem[{K{\"o}hler et~al.(2017)K{\"o}hler, Umlauf, Beyerlein, Winkler,
  Ziegler and Greven}]{kohler2017flexible}
\bibinfo{author}{K{\"o}hler, M.}, \bibinfo{author}{Umlauf, N.},
  \bibinfo{author}{Beyerlein, A.}, \bibinfo{author}{Winkler, C.},
  \bibinfo{author}{Ziegler, A.G.}, \bibinfo{author}{Greven, S.},
  \bibinfo{year}{2017}.
\newblock \bibinfo{title}{Flexible {B}ayesian additive joint models with an
  application to type~1 diabetes research}.
\newblock \bibinfo{journal}{Biometrical Journal} \bibinfo{volume}{59},
  \bibinfo{pages}{1144--1165}.
\newblock \DOIprefix\doi{10.1002/bimj.201600224}.
\bibitem[{Lang et~al.(2019)Lang, Mayr, Stauffer and
  Zeileis}]{lang2019bivariate}
\bibinfo{author}{Lang, M.N.}, \bibinfo{author}{Mayr, G.J.},
  \bibinfo{author}{Stauffer, R.}, \bibinfo{author}{Zeileis, A.},
  \bibinfo{year}{2019}.
\newblock \bibinfo{title}{Bivariate {G}aussian models for wind vectors in a
  distributional regression framework}.
\newblock \bibinfo{journal}{Advances in Statistical Climatology, Meteorology
  and Oceanography} \bibinfo{volume}{5}, \bibinfo{pages}{115--132}.
\newblock \DOIprefix\doi{10.5194/ascmo-5-115-2019}.
\bibitem[{Leutbecher and Palmer(2008)}]{leutbecher2008}
\bibinfo{author}{Leutbecher, M.}, \bibinfo{author}{Palmer, T.N.},
  \bibinfo{year}{2008}.
\newblock \bibinfo{title}{Ensemble forecasting}.
\newblock \bibinfo{journal}{Journal of Computational Physics}
  \bibinfo{volume}{227}, \bibinfo{pages}{3515--3539}.
\newblock \DOIprefix\doi{10.1016/j.jcp.2007.02.014}.
\bibitem[{Levina et~al.(2008)Levina, Rothman and Zhu}]{levina2008sparse}
\bibinfo{author}{Levina, E.}, \bibinfo{author}{Rothman, A.},
  \bibinfo{author}{Zhu, J.}, \bibinfo{year}{2008}.
\newblock \bibinfo{title}{Sparse estimation of large covariance matrices via a
  nested lasso penalty}.
\newblock \bibinfo{journal}{The Annals of Applied Statistics}
  \bibinfo{volume}{2}, \bibinfo{pages}{245--263}.
\newblock \DOIprefix\doi{10.1214/07-aoas139}.
\bibitem[{Mayr et~al.(2012)Mayr, Fenske, Hofner, Kneib and Schmid}]{mayr2012}
\bibinfo{author}{Mayr, A.}, \bibinfo{author}{Fenske, N.},
  \bibinfo{author}{Hofner, B.}, \bibinfo{author}{Kneib, T.},
  \bibinfo{author}{Schmid, M.}, \bibinfo{year}{2012}.
\newblock \bibinfo{title}{Generalized additive models for location, scale and
  shape for high dimensional data---a flexible approach based on boosting}.
\newblock \bibinfo{journal}{Journal of the Royal Statistical Society: Series C
  (Applied Statistics)} \bibinfo{volume}{61}, \bibinfo{pages}{403--427}.
\newblock \DOIprefix\doi{https://doi.org/10.1111/j.1467-9876.2011.01033.x}.
\bibitem[{Pan and Pan(2017)}]{pan2017jmcm}
\bibinfo{author}{Pan, J.}, \bibinfo{author}{Pan, Y.}, \bibinfo{year}{2017}.
\newblock \bibinfo{title}{{jmcm}: An {R} package for joint mean-covariance
  modeling of longitudinal data}.
\newblock \bibinfo{journal}{Journal of Statistical Software}
  \bibinfo{volume}{82}, \bibinfo{pages}{1--29}.
\newblock \DOIprefix\doi{10.18637/jss.v082.i09}.
\bibitem[{Pinson(2012)}]{pinson2012adaptive}
\bibinfo{author}{Pinson, P.}, \bibinfo{year}{2012}.
\newblock \bibinfo{title}{Adaptive calibration of $(u,v)$-wind ensemble
  forecasts}.
\newblock \bibinfo{journal}{Quarterly Journal of the Royal Meteorological
  Society} \bibinfo{volume}{138}, \bibinfo{pages}{1273--1284}.
\newblock \DOIprefix\doi{10.1002/qj.1873}.
\bibitem[{Pourahmadi(1999)}]{poura1999}
\bibinfo{author}{Pourahmadi, M.}, \bibinfo{year}{1999}.
\newblock \bibinfo{title}{Joint mean-covariance models with applications to
  longitudinal data: Unconstrained parameterisation}.
\newblock \bibinfo{journal}{Biometrika} \bibinfo{volume}{86},
  \bibinfo{pages}{677--690}.
\newblock \DOIprefix\doi{10.1093/biomet/86.3.677}.
\bibitem[{Pourahmadi(2000)}]{poura2000}
\bibinfo{author}{Pourahmadi, M.}, \bibinfo{year}{2000}.
\newblock \bibinfo{title}{Maximum likelihood estimation of generalised linear
  models for multivariate normal covariance matrix}.
\newblock \bibinfo{journal}{Biometrika} \bibinfo{volume}{87},
  \bibinfo{pages}{425--435}.
\newblock \DOIprefix\doi{10.1093/biomet/87.2.425}.
\bibitem[{Pourahmadi(2013)}]{poura2013high}
\bibinfo{author}{Pourahmadi, M.}, \bibinfo{year}{2013}.
\newblock \bibinfo{title}{High-Dimensional Covariance Estimation: With
  High-Dimensional Data}. volume \bibinfo{volume}{882}.
\newblock \bibinfo{publisher}{John Wiley \& Sons}.
\newblock \DOIprefix\doi{10.1002/9781118573617}.
\bibitem[{Rigby and Stasinopoulos(2005)}]{rigby2005generalized}
\bibinfo{author}{Rigby, R.A.}, \bibinfo{author}{Stasinopoulos, D.M.},
  \bibinfo{year}{2005}.
\newblock \bibinfo{title}{Generalized additive models for location, scale and
  shape}.
\newblock \bibinfo{journal}{Journal of the Royal Statistical Society C}
  \bibinfo{volume}{54}, \bibinfo{pages}{507--554}.
\newblock \DOIprefix\doi{10.1111/j.1467-9876.2005.00510.x}.
\bibitem[{Schefzik et~al.(2013)Schefzik, Thorarinsdottir and
  Gneiting}]{schefzik2013ecc}
\bibinfo{author}{Schefzik, R.}, \bibinfo{author}{Thorarinsdottir, T.L.},
  \bibinfo{author}{Gneiting, T.}, \bibinfo{year}{2013}.
\newblock \bibinfo{title}{Uncertainty quantification in complex simulation
  models using ensemble copula coupling}.
\newblock \bibinfo{journal}{Statistical Science} \bibinfo{volume}{28},
  \bibinfo{pages}{616--640}.
\newblock \DOIprefix\doi{10.1214/13-sts443}.
\bibitem[{Schoenach et~al.(2020)Schoenach, Simon and
  Mayr}]{schoenach2020vertical}
\bibinfo{author}{Schoenach, D.}, \bibinfo{author}{Simon, T.},
  \bibinfo{author}{Mayr, G.J.}, \bibinfo{year}{2020}.
\newblock \bibinfo{title}{Postprocessing ensemble forecasts of vertical
  temperature profiles}.
\newblock \bibinfo{journal}{Advances in Statistical Climatology, Meteorology
  and Oceanography} \bibinfo{volume}{6}, \bibinfo{pages}{45--60}.
\newblock \DOIprefix\doi{10.5194/ascmo-6-45-2020}.
\bibitem[{Schuhen et~al.(2012)Schuhen, Thorarinsdottir and
  Gneiting}]{schuhen2012ensemble}
\bibinfo{author}{Schuhen, N.}, \bibinfo{author}{Thorarinsdottir, T.L.},
  \bibinfo{author}{Gneiting, T.}, \bibinfo{year}{2012}.
\newblock \bibinfo{title}{Ensemble model output statistics for wind vectors}.
\newblock \bibinfo{journal}{Monthly Weather Review} \bibinfo{volume}{140},
  \bibinfo{pages}{3204--3219}.
\newblock \DOIprefix\doi{10.1175/mwr-d-12-00028.1}.
\bibitem[{Simon et~al.(2019)Simon, Mayr, Umlauf and Zeileis}]{simon2019nwp}
\bibinfo{author}{Simon, T.}, \bibinfo{author}{Mayr, G.J.},
  \bibinfo{author}{Umlauf, N.}, \bibinfo{author}{Zeileis, A.},
  \bibinfo{year}{2019}.
\newblock \bibinfo{title}{{NWP}-based lightning prediction using flexible count
  data regression}.
\newblock \bibinfo{journal}{Advances in Statistical Climatology, Meteorology
  and Oceanography} \bibinfo{volume}{5}, \bibinfo{pages}{1--16}.
\newblock \DOIprefix\doi{10.5194/ascmo-5-1-2019}.
\bibitem[{Stasinopoulos et~al.(2018)Stasinopoulos, Rigby and
  De~Bastiani}]{stasinopoulos2018gamlss}
\bibinfo{author}{Stasinopoulos, M.D.}, \bibinfo{author}{Rigby, R.A.},
  \bibinfo{author}{De~Bastiani, F.}, \bibinfo{year}{2018}.
\newblock \bibinfo{title}{{GAMLSS}: {A} distributional regression approach}.
\newblock \bibinfo{journal}{Statistical Modelling} \bibinfo{volume}{18},
  \bibinfo{pages}{248--273}.
\newblock \DOIprefix\doi{10.1177/1471082x18759144}.
\bibitem[{Umlauf et~al.(2021)Umlauf, Klein, Simon and
  Zeileis}]{Umlauf+Klein+Simon:2021}
\bibinfo{author}{Umlauf, N.}, \bibinfo{author}{Klein, N.},
  \bibinfo{author}{Simon, T.}, \bibinfo{author}{Zeileis, A.},
  \bibinfo{year}{2021}.
\newblock \bibinfo{title}{{bamlss}: A {L}ego toolbox for flexible {B}ayesian
  regression (and beyond)}.
\newblock \bibinfo{journal}{Journal of Statistical Software}
  \bibinfo{volume}{100}, \bibinfo{pages}{1--55}.
\newblock \bibinfo{note}{Forthcoming}.
\bibitem[{Umlauf et~al.(2018)Umlauf, Klein and Zeileis}]{umlauf2018bamlss}
\bibinfo{author}{Umlauf, N.}, \bibinfo{author}{Klein, N.},
  \bibinfo{author}{Zeileis, A.}, \bibinfo{year}{2018}.
\newblock \bibinfo{title}{{BAMLSS}: {Bayesian} additive models for location,
  scale, and shape (and beyond)}.
\newblock \bibinfo{journal}{Journal of Computational and Graphical Statistics}
  \bibinfo{volume}{27}, \bibinfo{pages}{612--627}.
\newblock \DOIprefix\doi{10.1080/10618600.2017.1407325}.
\bibitem[{Umlauf and Kneib(2018)}]{Umlauf+Kneib:2018}
\bibinfo{author}{Umlauf, N.}, \bibinfo{author}{Kneib, T.},
  \bibinfo{year}{2018}.
\newblock \bibinfo{title}{A primer on {B}ayesian distributional regression}.
\newblock \bibinfo{journal}{Statistical Modelling} \bibinfo{volume}{18},
  \bibinfo{pages}{219--247}.
\newblock \DOIprefix\doi{10.1177/1471082X18759140}.
\bibitem[{Wood(2003)}]{Wood:2003}
\bibinfo{author}{Wood, S.N.}, \bibinfo{year}{2003}.
\newblock \bibinfo{title}{Thin plate regression splines}.
\newblock \bibinfo{journal}{Journal of the Royal Statistical Society B}
  \bibinfo{volume}{65}, \bibinfo{pages}{95--114}.
\newblock \DOIprefix\doi{https://doi.org/10.1111/1467-9868.00374}.
\bibitem[{Wood(2017)}]{wood2017generalized}
\bibinfo{author}{Wood, S.N.}, \bibinfo{year}{2017}.
\newblock \bibinfo{title}{Generalized Additive Models: An Introduction with
  {R}}.
\newblock \bibinfo{edition}{2nd} ed., \bibinfo{publisher}{Chapman \& Hall/CRC},
  \bibinfo{address}{Boca Raton}.
\newblock \DOIprefix\doi{10.1201/9781315370279}.
\bibitem[{Worsnop et~al.(2018)Worsnop, Scheuerer, Hamill and
  Lundquist}]{worsnop2018generating}
\bibinfo{author}{Worsnop, R.P.}, \bibinfo{author}{Scheuerer, M.},
  \bibinfo{author}{Hamill, T.M.}, \bibinfo{author}{Lundquist, J.K.},
  \bibinfo{year}{2018}.
\newblock \bibinfo{title}{Generating wind power scenarios for probabilistic
  ramp event prediction using multivariate statistical post-processing}.
\newblock \bibinfo{journal}{Wind Energy Science} \bibinfo{volume}{3},
  \bibinfo{pages}{371--393}.
\newblock \DOIprefix\doi{10.5194/wes-3-371-2018}.
\bibitem[{Wu and Pourahmadi(2003)}]{wu2003}
\bibinfo{author}{Wu, W.B.}, \bibinfo{author}{Pourahmadi, M.},
  \bibinfo{year}{2003}.
\newblock \bibinfo{title}{Nonparametric estimation of large covariance matrices
  of longitudinal data}.
\newblock \bibinfo{journal}{Biometrika} \bibinfo{volume}{90},
  \bibinfo{pages}{831--844}.
\newblock \DOIprefix\doi{10.1093/biomet/90.4.831}.
\bibitem[{Zimmerman et~al.(1998)Zimmerman, N{\'u\~n}ez-Ant{\'o}n and
  El-Barmi}]{zimmerman1998}
\bibinfo{author}{Zimmerman, D.L.}, \bibinfo{author}{N{\'u\~n}ez-Ant{\'o}n, V.},
  \bibinfo{author}{El-Barmi, H.}, \bibinfo{year}{1998}.
\newblock \bibinfo{title}{Computational aspects of likelihood-based estimation
  of first-order antedependence models}.
\newblock \bibinfo{journal}{Journal of Statistical Computation and Simulation}
  \bibinfo{volume}{60}, \bibinfo{pages}{67--84}.
\newblock \DOIprefix\doi{10.1080/00949659808811872}.

\end{thebibliography}

\end{document}